\documentclass[onecolumn]{aastex631}

\usepackage{natbib}
\usepackage[version=4]{mhchem}
\usepackage{enumitem}
\usepackage{threeparttable}

\definecolor{background}{RGB}{252, 229, 241}
\definecolor{frame}{RGB}{225, 34, 138}

\newcommand{\RNum}[1]{\uppercase\expandafter{\romannumeral #1\relax}}

\DeclareSymbolFont{UPM}{U}{eur}{m}{n}
\DeclareMathSymbol{\umu}{0}{UPM}{"16}
\let\oldumu=\umu
\renewcommand\umu{\ifmmode\oldumu\else\math{\oldumu}\fi}

\usepackage{tikz}
\def\checkmark{\tikz\fill[scale=0.4](0,.35) -- (.25,0) -- (1,.7) -- (.25,.15) -- cycle;}

\shorttitle{Analyzing \ce{co2} with BARBIE \& KEN}
\shortauthors{Hagee et al.}

\begin{document}

\title{Bayesian Analysis for Remote Biosignature Identification on exoEarths (BARBIE) IV: Analyzing \ce{CO2} Detections in the Near-IR to Determine the Long-Wavelength Cut-off for the Habitable Worlds Observatory Coronagraph}

\author[0000-0001-7441-241X]{Celeste Hagee}
\affiliation{NASA Goddard Space Flight Center, 8800 Greenbelt Road, Greenbelt, MD 20771, USA}
\affiliation{Sellers Exoplanets Environment Collaboration, 8800 Greenbelt Road, Greenbelt, MD 20771, USA}
\affiliation{Southeastern Universities Research Association, 1201 New York Ave. NW, Suite 430
Washington, DC 20005 USA}

\author[0000-0001-8079-1882]{Natasha Latouf}
\altaffiliation{NSF Graduate Research Fellow, 2415 Eisenhower Ave, Alexandria, VA 22314}
\affiliation{Department of Physics and Astronomy, George Mason University, 4400 University Drive MS 3F3, Fairfax, VA 22030, USA}
\affiliation{NASA Goddard Space Flight Center, 8800 Greenbelt Road, Greenbelt, MD 20771, USA}
\affiliation{Sellers Exoplanets Environment Collaboration, 8800 Greenbelt Road, Greenbelt, MD 20771, USA}

\author[0000-0002-8119-3355]{Avi M. Mandell}
\affiliation{NASA Goddard Space Flight Center, 8800 Greenbelt Road, Greenbelt, MD 20771, USA}
\affiliation{Sellers Exoplanets Environment Collaboration, 8800 Greenbelt Road, Greenbelt, MD 20771, USA}

\author[0000-0002-9338-8600]{Michael D. Himes}
\affiliation{Morgan State University, 1700 E Cold Spring Lane, Baltimore, MD 21251, USA}
\affiliation{NASA Goddard Space Flight Center, 8800 Greenbelt Road, Greenbelt, MD 20771, USA}

\author[0000-0001-7912-6519]{Michael Dane Moore}
\affiliation{NASA Goddard Space Flight Center, 8800 Greenbelt Road, Greenbelt, MD 20771, USA}
\affiliation{Business Integra, Inc., Bethesda, MD, USA.}
\affiliation{Sellers Exoplanets Environment Collaboration, 8800 Greenbelt Road, Greenbelt, MD 20771, USA}


\author[0000-0002-2662-5776]{Geronimo L. Villanueva}
\affiliation{NASA Goddard Space Flight Center, 8800 Greenbelt Road, Greenbelt, MD 20771, USA}
\affiliation{Sellers Exoplanets Environment Collaboration, 8800 Greenbelt Road, Greenbelt, MD 20771, USA}


\correspondingauthor{Celeste Hagee}
\email{celeste.hagee@email.ucr.edu}

\begin{abstract}
    We present our analysis of how the detectability of carbon dioxide (\ce{CO2}) on an Earth-like planet varies with respect to signal-to-noise ratio (SNR), wavelength, and molecular abundance. Using the Bayesian Analysis for Remote Biosignature Identification on exoEarths (BARBIE) methodology, we can inform the optimal long-wavelength cut-off for the future Habitable Worlds Observatory (HWO) coronagraph. We test 25 evenly-spaced 20\% bandpasses between 0.8-2.0$\mu$m, and simulate data spanning a range of SNRs and molecular abundance to analyze the relationship between wavelength and detectability for different planetary archetypes. We examine abundance levels from varying Earth epochs and a Venus-like archetype to investigate how detectability would change throughout the evolution of a rocky planet. Here, we present our results on the planetary conditions and technological requirements to strongly detect \ce{CO2}. In addition, we analyze the degeneracy of \ce{CO2} with carbon monoxide (\ce{CO}), methane (\ce{CH4}), and water (\ce{H2O}). We determine that any abundance of \ce{CO} does not achieve strong detections and that \ce{CH4} and \ce{H2O} play a pivotal role in the ability to detect \ce{CO2}. We conclude that the optimal long-wavelength cut-off for the Habitable Worlds Observatory coronagraph should be 1.68$\mu$m.
\end{abstract}

\section{Introduction}
Since the first discovery of an exoplanet around a main-sequence star in 1995 \citep{mayor95}, more than 6,000 exoplanets have been discovered \citep{nasa-exoplanet-archive}. Astronomers are now looking towards characterizing these exoplanets, especially small, terrestrial planets, to tell us more about their formation, evolution, and composition. However, in order to constrain the properties of potentially Earth-like exoplanets around Sun-like stars, future telescopes will need to use high-contrast direct imaging, which requires a highly effective starlight suppression system such as a high-contrast coronagraph or starshade \citep{roberge18, luvoir, gaudi20}. Based on the recommendations of the \citet{decadal}, NASA began developing a future space telescope that could accomplish this goal; this concept is now called the Habitable Worlds Observatory (HWO). HWO will be designed to have the technological imaging capability to detect and characterize Earth-like planets and identify signs of planetary habitability and possibly biology. Naturally, one of the key instruments currently in active development for HWO is the coronagraph instrument, nominally planned to have the capability to observe in $\sim$20\% bandpasses across the UV, visible, and NIR wavelength regimes \citep{luvoir,roser22}.  

A yet-undetermined parameter for the coronagraph instrument is the long-wavelength cut-off. One of the driving factors for constraining the long-wavelength cut-off in the NIR is the telescope temperature. At higher telescope temperatures, a higher background flux will contribute to a higher overall photon noise, impacting the exposure time necessary to reach a required SNR. Two strategies can help avoid the impact of a higher thermal background on spectral characterization: targeting spectral features at shorter wavelengths, where absorption features are weaker but thermal background is minimal, or cooling the telescope to enable low-background noise measurements of deeper spectral features at longer wavelengths. 

\begin{figure}[hbt!]
\centering
\includegraphics[scale=0.5]{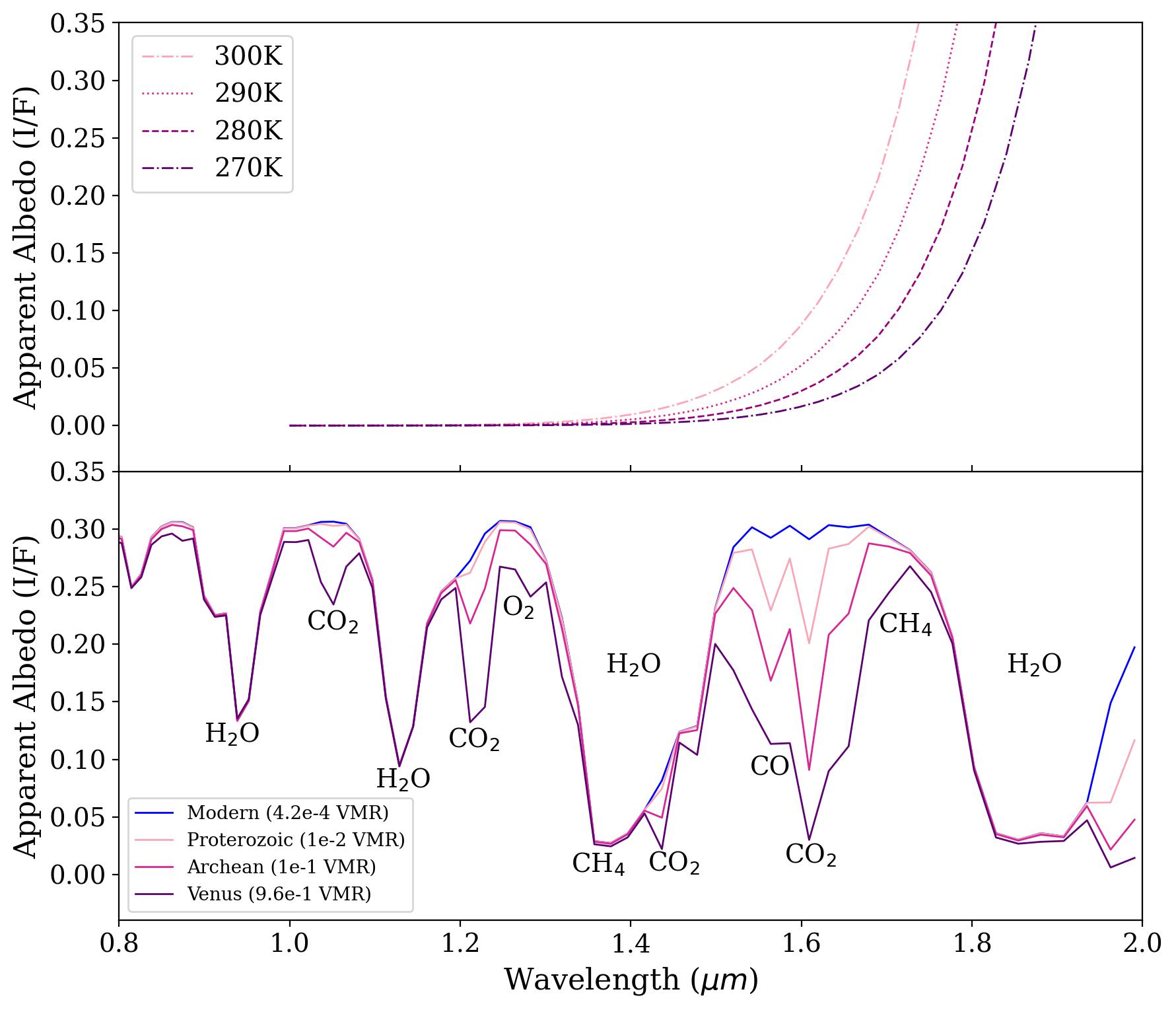}
\caption{Illustrates the balancing act of keeping thermal noise minimal while maximizing the detectability of spectral features like \ce{CO2} and \ce{CO}. The top plot shows the relative noise based on varying telescope temperature between 0.8-2.0$\mu$m. The bottom plot shows spectral features from 0.8-2.0$\mu$m of \ce{H2O}, \ce{CH4}, \ce{O2}, \ce{CO}, and \ce{CO2}. We plot a modern-Earth spectra using the Planetary Spectrum Generator (PSG) with four varying \ce{CO2} abundances: Modern (4.2$\times10^{-4}$ VMR) in dark blue, Proterozoic-like (1$\times10^{-2}$ VMR) in light pink, Archean-like (1$\times10^{-1}$ VMR) in magenta, and Venus-like (9.6$\times10^{-1}$ VMR) in purple. The figure plots wavelength ($\mu$m) on the x-axis and apparent albedo (I/F) on the y-axis.}
\label{fig:intro}
\end{figure}

Figure~\ref{fig:intro} illustrates this trade-off. The top panel shows how noise from the blackbody emission of the telescope, henceforth called thermal noise, fluctuates with respect to four varying temperatures (270K in dark purple, 280K in purple, 290K in deep pink, and 300K in light pink). The bottom panel illustrates how spectral features are impacted by wavelength and molecular abundance by plotting four modern-Earth spectra with Modern, Proterozoic-like, Archean-like, and Venus-like \ce{CO2} abundances with labeled water (\ce{H2O}), methane (\ce{CH4}), oxygen (\ce{O2}), carbon monoxide (\ce{CO}), and carbon dioxide (\ce{CO2}) features over 0.8-2.0$\mu$m.

It is clear that for wavelengths past $\sim$1.3$\mu$m, higher telescope temperatures produce more noise. However, cooling a massive space telescope below room temperature ($\sim$300K) increases the overall cost dramatically, due to the need for fabrication and testing facilities operating at these temperatures. Additionally, cooling a telescope below room temperature has shown decreased stability in dynamic damping and mirror materials and would require abandoning the current coronagraph technology developments \citep{roomtemp}. Cooling the telescope could be reevaluated in the future if advancements such as new cryogenic technologies are made. However, we move forward with this study based on the current baseline of having a telescope operating at room temperature. Thus, our goal is to determine whether exoplanet characterization is possible with a room temperature telescope (i.e., below 1.7$\mu$m). This will enable us to find the optimal long-wavelength cut-off that provides the opportunities to maximize our spectral characterization while minimizing additional costs.


At NIR wavelengths, many molecules of interest have absorption features, as shown in the bottom panel of Figure~\ref{fig:intro}. In particular, we can visually see that \ce{H2O}, \ce{CH4}, and \ce{O2} all have deep features at shorter NIR wavelengths, and thus are not the science drivers for the coronagraph long-wavelength cut-off. However, \ce{CO2} and \ce{CO} have their most absorbent features at wavelengths longer than 1.4$\mu$m, with relatively weak features (or no features) at shorter wavelengths. To illustrate this further, \ce{CO2} and \ce{CO} spectra at three varying abundances each - instead of a full modern-Earth spectra - are plotted in Figure~\ref{fig:co}. It is evident that \ce{CO2} features are significantly deeper at longer wavelengths. The only apparent \ce{CO} feature is at 1.58$\mu$m, and even at the highest abundance - 1$\times10^{-2}$ VMR - the absorption depth does not compare with \ce{CO2} features at the same wavelength. Thus, the main focus of this study is to determine the optimal bandpass that \ce{CO2} can be strongly detected to constrain the long-wavelength cut-off for the HWO coronagraph.



We used the established Bayesian Analysis for Remote Biosignature Identification on exoEarths (BARBIE) project methodology (\cite{barbie1,barbie2,barbie3}, hereafter BARBIE1, BARBIE2, BARBIE3, respectively) to understand the detectability of \ce{CO2} and the molecules that could impact it such as \ce{CO}, \ce{H2O}, and \ce{CH4}. Following the BARBIE methodology, our goal is determine the optimal single bandpass across the full 0.8-2.0$\mu$m spectral range that would provide the best opportunity to achieve a strong \ce{CO2} detection in a single observation. This would enable a solid detection even under the scenario where HWO only performs spectroscopy with a single coronagraphic bandpass. Under a scenario with multiple parallel coronagraphic channels, this enables us to identify which bandpasses can cover a wide range of molecules. Thus, each channel can be optimized to obtain spectral information from as many molecules in the exoplanet's atmosphere as possible. This research analyzes unknown Earth-like exoplanet atmosphere compositions by testing a wide range of \ce{CO2}, \ce{H2O}, and \ce{CH4} abundances throughout the history of an Earth-like planet. Additionally, we get a deeper understanding of how overlapping \ce{CO2}, \ce{CO}, \ce{H2O}, and \ce{CH4} spectral features impact each other and their detectability. Combined with previous BARBIE studies, which analyzed the detectability of \ce{O2}, \ce{O3}, \ce{H2O}, and \ce{CH4} at shorter wavelengths, we will be able to determine the optimal long-wavelength cut-off that ensures we achieve strong detections for all biosignatures of interest with minimal engineering complications at a reasonable cost for the HWO coronagraph. In Section~\ref{sec:method}, we provide an overview of our methodology, including descriptions of the KEN grids. In Section~\ref{sec:results}, we present our results from the simulations of \ce{CO2} on Earth-like and Venus-like planetary archetypes using two different grids: L-KEN and B-Ken. Furthermore, we discuss our analysis of \ce{CO2}-\ce{H2O} and \ce{CO2}-\ce{CH4} degeneracies and how our results inform the optimal HWO coronagraph long-wavelength cut-off. Lastly, we analyze how four planetary archetypes, with combinations of low and high abundances of \ce{CH4} and \ce{H2O}, would perform given our long-wavelength cut-off. In Section~\ref{sec:end}, we present our conclusions and discuss our next steps in the BARBIE world.




\section{Methodology}
\label{sec:method}

\subsection{BARBIE Methodology}

We follow the same methodology as BARBIE1, BARBIE2 \& BARBIE3 to analyze the detectability of \ce{CO2} under different assumptions for abundance and wavelength. In this section, we summarize the main steps of our analysis. For additional information on the details of model grid development, validation, and spectral retrieval methodology, please refer to the previous BARBIE studies.

\begin{table}
\centering
\caption{Atmospheric parameters for a modern-Earth twin for our fiducial data spectrum following \cite{feng18}}
\begin{tabular}{|c|c|}
\hline
    Parameter & Value \\
\hline
    \ce{H2O} Isotropic VMR & $3\times10^{-3}$ \\ 
    \ce{CH4} Isotropic VMR & $1.65\times10^{-6}$ \\ 
    \ce{CO2} Isotropic VMR & $3.8\times10^{-4}$ \\
\hline
    Constant temperature profile & 250K \\
    Surface Albedo $(\mathrm{A_s})$ & 0.3 \\
    Pressure $(\mathrm{P_0})$ & 1 bar \\
    Fixed planetary radius at $\mathrm{R_p}$ & 1 $\mathrm{R_\Earth}$ \\
\hline
\end{tabular}
\label{tab:fidiucial}
\end{table}


First, following \cite{feng18}, we set a modern-Earth twin with an isotropic atmosphere for our fiducial data spectrum as seen in Table~\ref{tab:fidiucial}. We also compiled an array of literature-sourced abundances for \ce{CO2} from various epochs of Earth's history \citep{MAHIEUX2023115713, kaltenegger2007, cenozoic, modern}, as shown in Table~\ref{tab:abundances}, in order to simulate Earth's spectrum at different snapshots of time, enabling the investigation of how detectability varies across a physically-motivated range of abundances. BARBIE is a 1D model that utilizes the Planetary Spectrum Generator (PSG) \citep[PSG;][]{Villanueva_2018} for the spectral simulations. PSG is a publicly accessible radiative transfer modeling tool that generates synthetic planetary spectral data across UV to Radio wavelengths for various target and observation assumptions. Molecular opacities are obtained from PSG which utilizes HITRAN \citep{gordon20}. The spectra are simulated with an atmosphere that is 50\% cloudy and 50\% clear.

The PSG framework for target and observatory parameterization has also been used for the related Bayesian inference tool called PSGnest\footnote{https://psg.gsfc.nasa.gov/apps/psgnest.php}. PSGnest utilizes a custom implementation of the MultiNest nested sampling routine to enable Bayesian nested sampling retrievals using pre-computed grids of PSG-derived spectra. The PSGnest results produce several outputs: the maximum-likelihood values, the averages from the posterior distributions, uncertainties, and log-evidence (logZ) \citep{Villanueva_2018}. For each retrieval, we calculate the log-Bayes factor \citep[$\mathrm{lnB}$;][]{benneke13} by subtracting the Bayesian log-evidence for a retrieval without the molecule from the Bayesian log-evidence for a retrieval with the molecule. Thus, $\mathrm{lnB}$ directly compares fits by examining if an improved fit occurs with the molecule included, and addresses the likelihood of the presence of the molecular species and its detectability. We shift our definition of detection strength from the original \citet{benneke13} as detailed in Table~\ref{tab:lnB}. If you are interested in the relationship between the log-Bayes Factor and the sigma value, please refer to Table 2 in \cite{benneke13}.

$\mathrm{lnB}$ directly investigates the presence or absence of the molecule in the atmosphere and calculates which scenario results in a better fit. $\mathrm{lnB}$ is only calculated for gaseous components since those factors can be absent. We also calculate the upper limit, median value, and lower limit of the 68\% credible region \citep{harrington22}. The 68\% credible region includes the true value 68\% of the time and informs us of the range of molecular abundances that are consistent with our data. 


\begin{table}
\centering
\caption{Detection strength definitions}
\begin{tabular}{|c|c|}
\hline
    Criterion & Detection Strength \\
\hline
    $\mathrm{lnB}$ $\mathrm{<}$ 2.5 & unconstrained \\ 
    $2.5 \leq$ $\mathrm{lnB}$ $\mathrm{<}$ 5 & weak \\ 
   $\mathrm{lnB}$ $\geq 5$ & strong \\
\hline
\end{tabular}
\label{tab:lnB}
\end{table}



The abundance of \ce{CO2} has drastically varied over the course of Earth's history \citep{kaltenegger2007}. In a modern Earth-like planet, there are very limited amounts of \ce{CO2} due to natural processes that remove it from the atmosphere. For example, photosynthesis takes \ce{CO2} out of the air and produces \ce{O2} (as described in detail in \cite{photosythesis}); similarly, \ce{CO2} is absorbed by chemical weathering - which transforms silicate rocks into carbonate rocks - and the weathering of carbonate minerals \citep{weathering}. Models of the Archean epoch suggest that \ce{CO2} was much higher than now \citep{kaltenegger2007}. Processes like these emphasize the need to study a wide range of \ce{CO2} abundances. Determining the detectability strength of \ce{CO2} given a variety of abundances will allow for a better understanding of the possibility of molecular detection on Earth-like planets with HWO. 

In our study, we initially used modern-Earth abundances of \ce{H2O}, \ce{CH4}, and \ce{N2}. We varied the abundance of \ce{CO2} through five Earth-era abundances motivated by prior studies \citep{kaltenegger2007, cenozoic, modern}. It is important to note that we tested two additional abundances that do not connect directly to Earth epochs to ensure a broad range of possible values without large leaps in abundance space, including a Venus-like abundance \citep{MAHIEUX2023115713} and a 'bridging' abundance to examine the region in our abundance space between the Cenozoic and Proterozoic abundances. All \ce{CO2} abundance values are shown in Table~\ref{tab:abundances}. 

\begin{table}
\centering
\caption{Varying \ce{CO2} abundances and their corresponding atmospheric archetypes}
\begin{threeparttable}
\begin{tabular}{|c|c|}
\hline
    \ce{CO2} Abundance (VMR) & Atmospheric Archetypes \\
\hline
    9.6$\times10^{-1}$ & Venus-like\tnote{a} \\
    1$\times10^{-1}$ & Archean\tnote{b} \\
    1$\times10^{-2}$ & Proteroizoic\tnote{b} \\
    1$\times10^{-3}$ & Bridging \\
    3.65$\times10^{-4}$ & Paleozoic\tnote{b} \\
    7.2$\times10^{-4}$ & Cenozoic\tnote{c} \\
    4.2$\times10^{-4}$ & Modern\tnote{d} \\
\hline
\end{tabular}
\begin{tablenotes}
\item[a] \cite{MAHIEUX2023115713}
\item[b] \cite{kaltenegger2007}
\item[c] \cite{cenozoic}
\item[d] \cite{modern}
\end{tablenotes}
\label{tab:abundances}
\end{threeparttable}
\end{table}


Additionally, due to the prominent overlapping spectral features of \ce{CO2}, \ce{CH4}, and \ce{H2O} - which increases the difficulty in distinguishing the differences between molecules - we investigate the effect \ce{CO2} has on the detectability of \ce{H2O} and \ce{CH4} and vice versa by testing a range of combined values. \ce{CO2}-\ce{H2O} test values are shown in the first and second columns of Table~\ref{tab:all_abundances}. \ce{CO2}-\ce{CH4} test values are shown in the first and third columns of Table~\ref{tab:all_abundances}.

\begin{table}
\centering
\caption{Varying \ce{CO2}, \ce{H2O}, and \ce{CH4} abundances used in the dual-detection simulations}
\begin{tabular}{|c|c|c|}
\hline
    \ce{CO2} Abundance (VMR) & \ce{H2O} Abundance (VMR) & \ce{CH4} Abundance (VMR)\\
\hline
    9.6$\times10^{-1}$ & 1$\times10^{-2}$ &7$\times10^{-3}$ \\
    1$\times10^{-1}$ & 3$\times10^{-3}$ &3$\times10^{-3}$ \\
    1$\times10^{-2}$ & 1$\times10^{-3}$ &1$\times10^{-3}$ \\
    1$\times10^{-3}$ & 3$\times10^{-4}$ &3$\times10^{-4}$ \\
    7.2$\times10^{-4}$ & 1$\times10^{-4}$ &1$\times10^{-4}$ \\
\hline
\end{tabular}
\label{tab:all_abundances}
\end{table}

We constrain our wavelength regime to wavelengths between 0.8-2.0$\mu$m, as there are no prominent \ce{CO2} features at shorter wavelengths. Our grid is binned from a native resolving power of 500 to 70, following \citet{luvoir} for NIR resolving power. For the simulations, we split the spectrum into 25 evenly spaced bandpasses of 20\% widths to mimic a coronagraphic observation. We also vary the SNRs from 3-20 to understand at what SNRs different \ce{CO2} abundances would be detectable.

\subsection{KEN Grids}

All BARBIE studies use grid-based retrievals. To ensure efficient production of new optimized spectral grids for grid-based retrievals, \citet{himes24} developed the Python package, Gridder, which is a generalized grid-building scheme based on the methodology of \citet{susemiehl23}. Gridder produces arbitrary spectral grids using PSG. The Gridder parameter structure is very customizable, allowing any thermal profile and atmospheric chemistry in PSG to be used as a parameter in the grid. 

Using Gridder, BARBIE3 \citep{barbie3} developed and validated a new set of spectral grids called the KEN (no acronym - they're just KEN). These grids cover a larger wavelength range (0.2 - 2.0 $\mu$m) at a native resolving power of R = 500, and have the same three base parameters: surface pressure ($\mathrm{P_0}$), surface albedo ($\mathrm{A_s}$), and gravity ($\mathrm{g}$). KEN consists of four grids, each with a set of four different molecules, in addition to the three base parameters as seen in Table~\ref{tab:ken}.


\begin{table}[hbt!]
\caption{KEN Grids}
\centering
\begin{tabular}{|c|c|c|c|c|c|c|c|c|c|}
\hline
    Grid Name & \ce{H2O} & \ce{CO2} & \ce{CO} & \ce{O3} & \ce{O2} & \ce{CH4} & \ce{SO2} & \ce{N2O} & \ce{N2} \\
\hline
    Merman Grid (M-KEN) & \checkmark &  &  & \checkmark & \checkmark &  &  &  & \checkmark \\
\hline
    Beach Grid (B-KEN) & \checkmark & \checkmark &  &  &  & \checkmark &  &  & \checkmark \\
\hline
    Allan Grid (A-KEN) &  &  &  & \checkmark &  &  & \checkmark & \checkmark & \checkmark \\
\hline
    Lifeguard Grid (L-KEN) & \checkmark & \checkmark & \checkmark &  &  &  &  &  & \checkmark \\
\hline
\end{tabular}
\label{tab:ken}
\end{table}


When choosing a grid to use, we consider the molecule of interest and the relationship it has with other molecules. Thus, we used the L-KEN and B-KEN grids to investigate \ce{CO2} and its relationships with \ce{CO} (L-KEN), \ce{H2O} (B-KEN), and \ce{CH4} (B-KEN) in the NIR. For more information on the KEN grids and grid building, please refer to \cite{barbie3}.

\section{Results}
\label{sec:results}

\subsection[CO2-CO]{\ce{CO2}-\ce{CO} Detectability}
\label{sec:co2-co}



\begin{figure}[hbt!]
\centering
\includegraphics[scale=0.5]{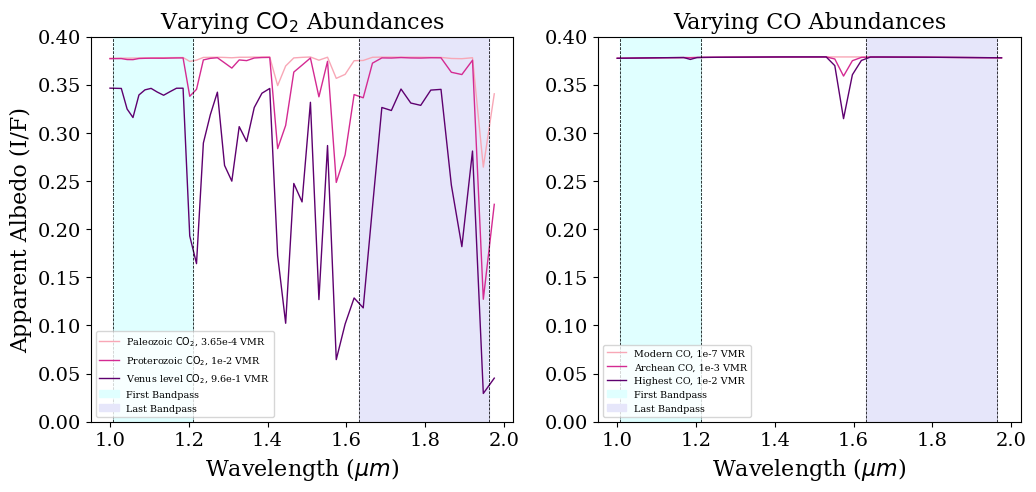}
\caption{\ce{CO2} and \ce{CO} spectral features at varying abundances from 0.8-2.0$\mu$m. It plots wavelength ($\mu$m) on the x-axis and apparent albedo (I/F) on the y-axis. We plot three \ce{CO2} abundances from Table~\ref{tab:abundances}: Paleozoic-like (3.65$\times10^{-4}$ VMR) in light pink, Proterozoic-like (1$\times10^{-2}$ VMR) in magenta, and a Venus-like (9.6$\times10^{-1}$ VMR) in purple. We plot three \ce{CO} abundances: Modern (1$\times10^{-7}$ VMR - \cite{co-modern}) in light pink, Archean-like (1$\times10^{-3}$ VMR - \cite{co-archean}) in magenta, and the highest abundance (1$\times10^{-2}$ VMR - filler abundance) in purple. The first and last bandpasses are shaded in light blue and light purple, respectively.}
\label{fig:co}
\end{figure}

\begin{figure}[hbt!]
\centering
\includegraphics[scale=0.31]{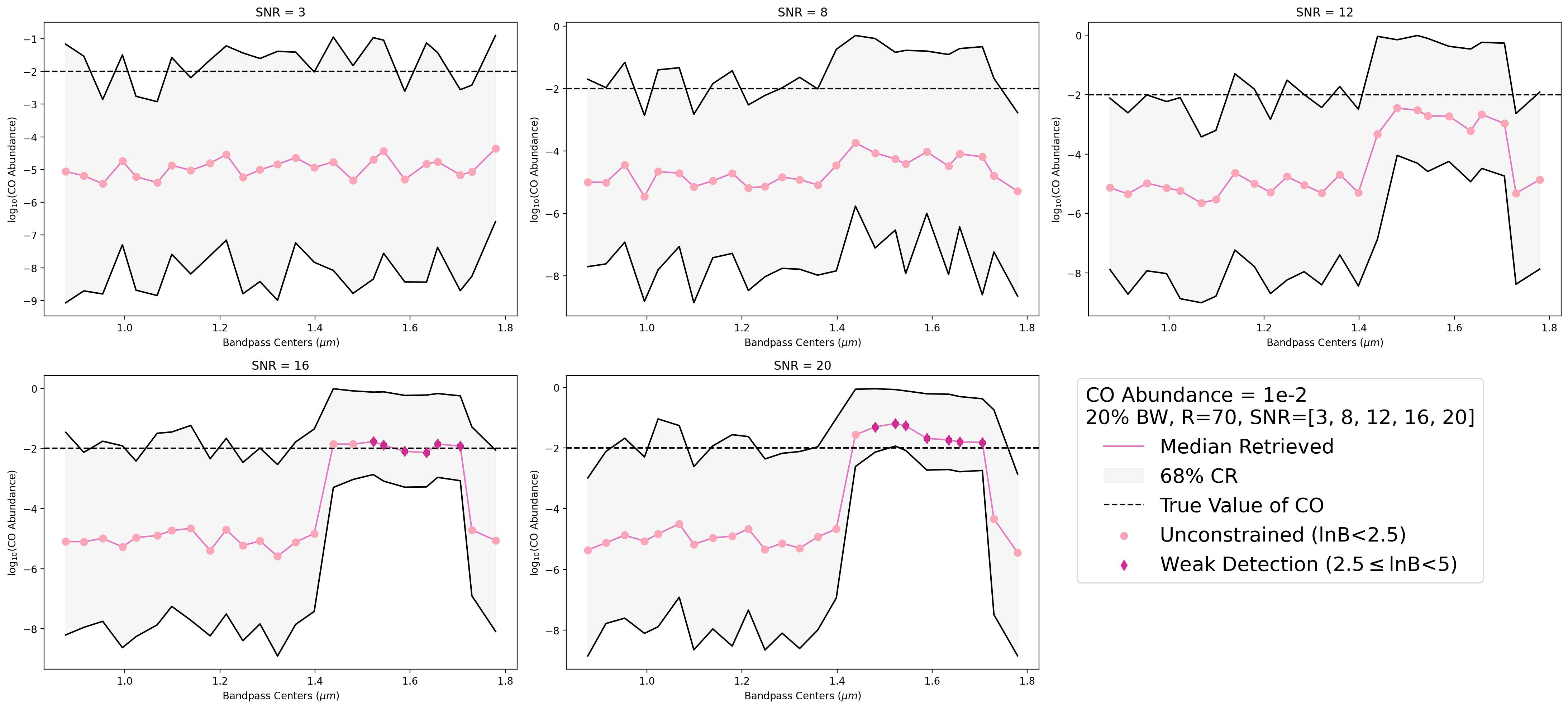}
\caption{Strength of \ce{CO} detections at SNRs of 3, 8, 12, 16 \& 20 with the highest \ce{CO} abundance, 1$\times10^{-2}$ VMR, throughout the wavelength range of 0.8-2.0$\mu$m. Light pink circles portray unconstrained detections, magenta diamonds portray weak detections, and there are no strong detections. $\mathrm{lnB}$ classification follows the values in Table~\ref{tab:lnB}. The 68\% credible region is shaded in grey. The true value of the \ce{CO} abundance is marked by the horizontal dashed black line.}
\label{fig:co_detectability}
\end{figure}

We initially used L-KEN to study the detectability of \ce{CO2} and \ce{CO} and found that with the highest \ce{CO} abundance, 1$\times10^{-2}$ VMR, no strong \ce{CO} detections are achievable. This is shown in Figure~\ref{fig:co_detectability}; with a SNR$=$20, only a weak detection is achievable between 1.48-1.71$\mu$m. Additionally, Figure~\ref{fig:co} emphasizes the drastic difference in feature size between varying \ce{CO2} and \ce{CO} features. Comparing spectral features of a Proterozoic-like \ce{CO2} abundance and the highest \ce{CO} abundance - both 1$\times10^{-2}$ VMR - at 1.58$\mu$m, the \ce{CO2} feature diminishes the apparent albedo by 0.3 whereas the \ce{CO} feature diminishes the apparent albedo by 0.05. Since they are overlapping, the \ce{CO} feature is overpowered by the larger \ce{CO2} feature, emphasizing the difficultly of strongly detecting \ce{CO}. Due to this difference in spectral feature depth, \ce{CO} would have an insignificant effect on the ability to detect \ce{CO2}. Therefore, for the rest of the study, we use B-KEN to study the impact of \ce{H2O} and \ce{CH4} on the detectability of \ce{CO2}, since these species have distinct and significant overlapping features in our wavelength range (shown in the bottom panel of Figure~\ref{fig:intro}) and would more likely impact the detectability of \ce{CO2} at various abundances.


\subsection[CO2-Modern]{\ce{CO2} Detectability with Modern \ce{H2O} \& \ce{CH4} Abundances}
\label{sec:co2-alone}

In this section, we present our results for \ce{CO2} detectability with modern abundances of \ce{H2O} and \ce{CH4}. We note that Modern, Paleozoic-like, Cenozoic-like, and Bridging Earth \ce{CO2} abundances do not produce strong \ce{CO2} detections at any wavelength, abundance, or SNR with modern levels of \ce{CH4} and \ce{H2O}. Thus, our discussion will focus on the three highest \ce{CO2} abundances that produce strong \ce{CO2} detections at modern \ce{H2O} and \ce{CH4} abundances: Proterozoic-like, Archean-like, and Venus-like.

\begin{figure*}
\centering
\gridline{\fig{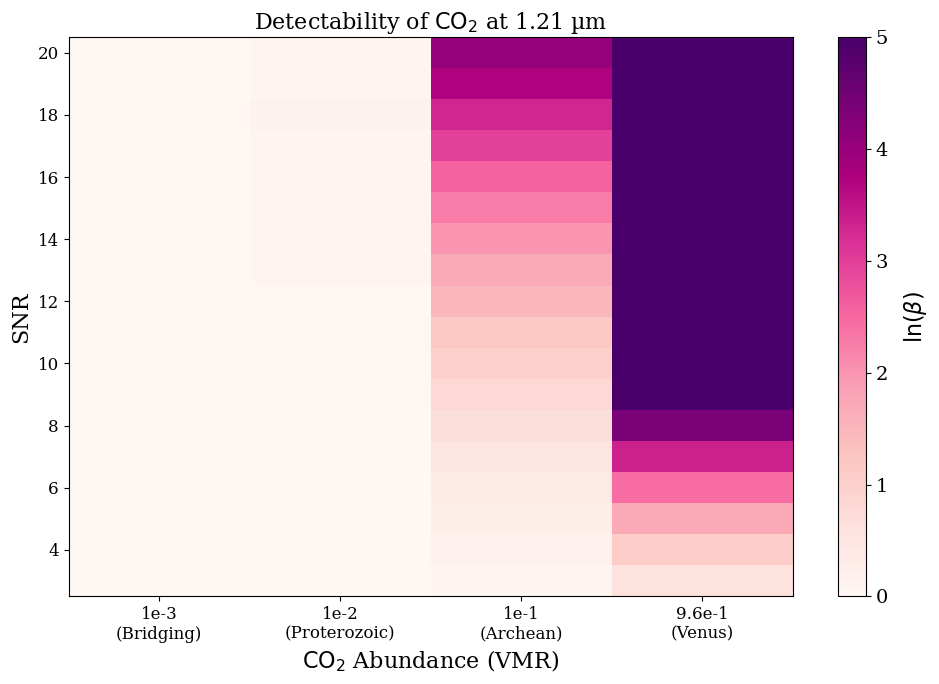}{0.4\textwidth}{\vspace{-0.35cm}(a) 1.21$\mu$m}
    \fig{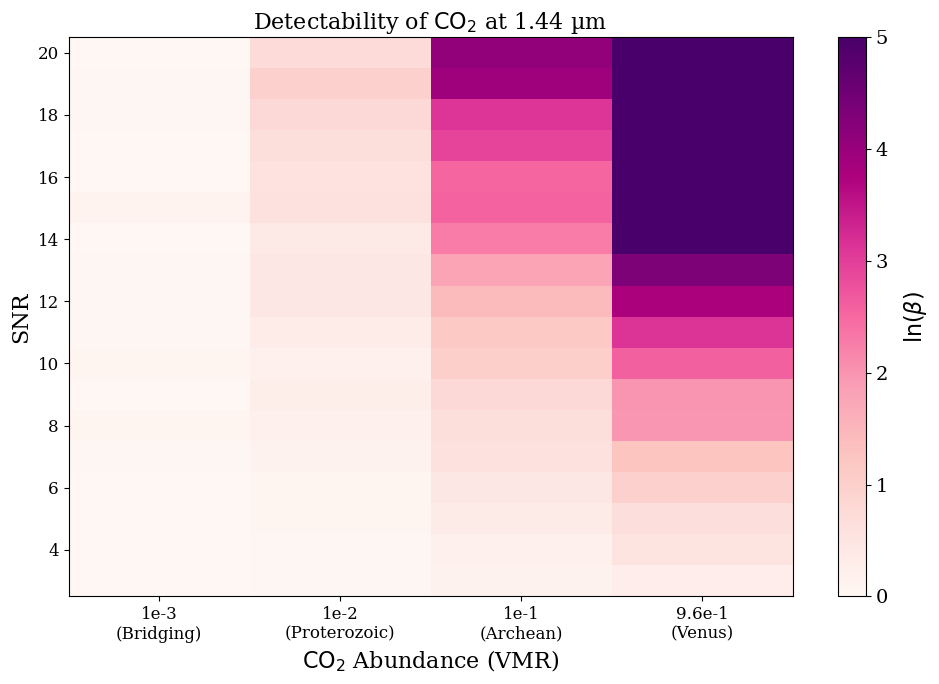}{0.4\textwidth}{\vspace{-0.35cm}(b) 1.44$\mu$m}}
\gridline{\fig{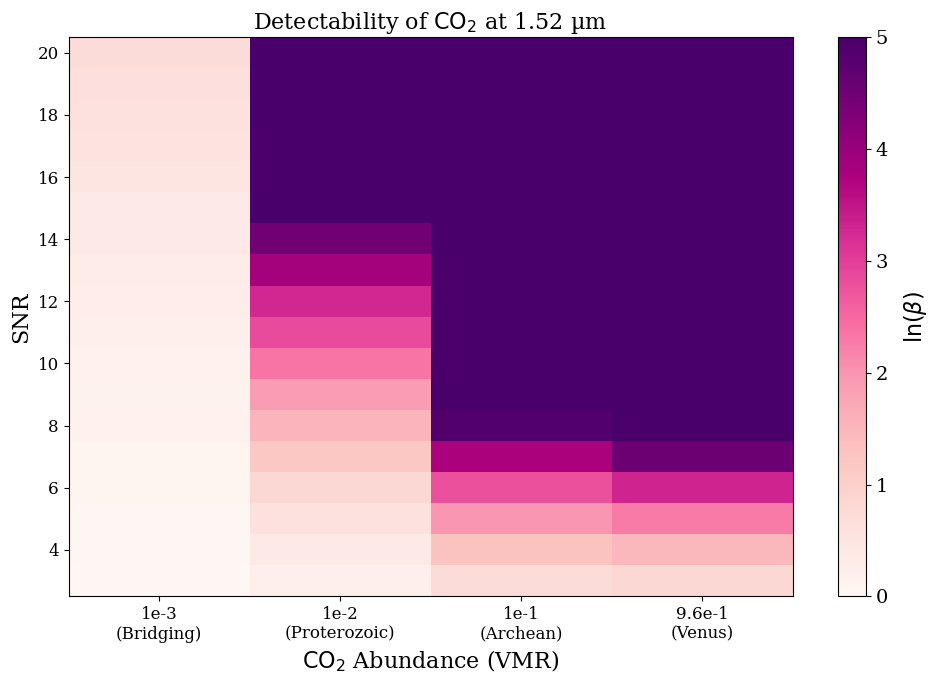}{0.4\textwidth}{\vspace{-0.35cm}(c) 1.52$\mu$m}
    \fig{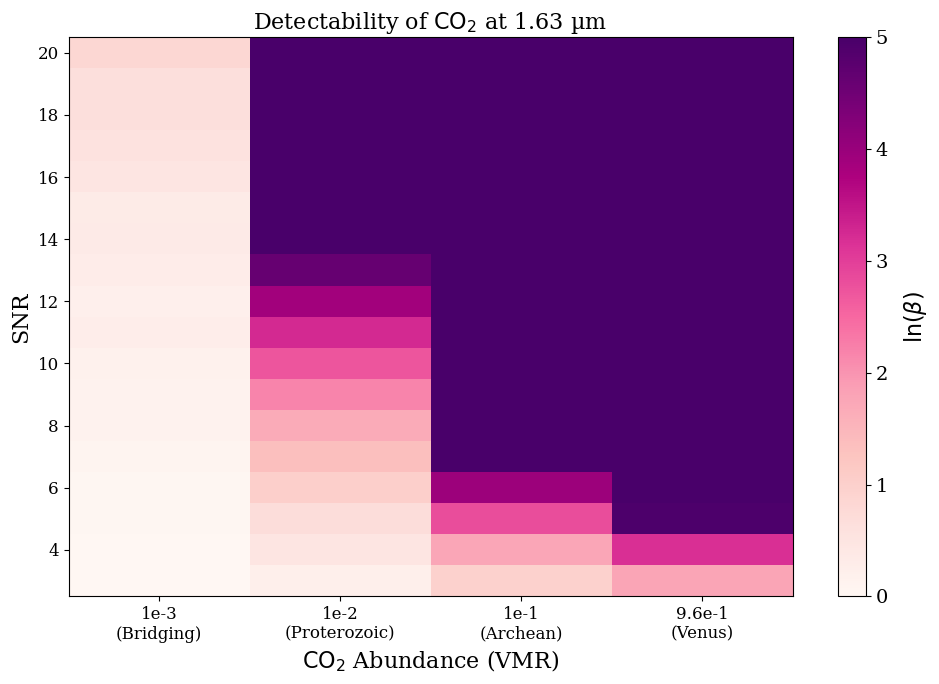}{0.4\textwidth}{\vspace{-0.35cm}(d) 1.63$\mu$m}}
\caption{Strength of \ce{CO2} detections at (a) 1.21 $\mu$m, (b) 1.44$\mu$m, (c) 1.52$\mu$m, and (d) 1.63$\mu$m with respect to \ce{CO2} abundance and SNR assuming a 20\% bandpass. Columns on the x-axis are different \ce{CO2} abundances; Bridging, Proterozoic-like, Archean-like, and Venus-like, from left to right. SNR values are on the y-axis. The strength of detection is shown for each combination of \ce{CO2} abundance and SNR. As indicated by the color bar on the right side of each heat map, the darker the purple is, the stronger the detection is, and the lighter the pink is, the more unconstrained the detection is, following the same $\mathrm{lnB}$ classifications defined in Table~\ref{tab:lnB}.}
\label{fig:heatmaps}
\end{figure*}

From Figure~\ref{fig:co}, we chose four bandpass centers - 1.21$\mu$m, 1.44$\mu$m, 1.52$\mu$m, and 1.63$\mu$m - that align with the deepest \ce{CO2} spectral features shortwards of 1.95$\mu$m. At each bandpass center, we present the detectability strength of the four highest \ce{CO2} abundances as a function of SNR and abundance in Figure~\ref{fig:heatmaps}. The strength of detection is shown for each combination of \ce{CO2} abundance and SNR, where the darker the purple, the stronger the detection, and the lighter the pink, the more unconstrained the detection. Figure~\ref{fig:heatmaps} provides clear insight into how wavelength, abundance, and SNR impact \ce{CO2} detectability.

In Figure~\ref{fig:heatmaps}a, at 1.21$\mu$m, a Venus-like \ce{CO2} abundance is necessary to achieve a strong detection starting at SNR$=9$. In Figure~\ref{fig:heatmaps}b, at 1.44$\mu$m, a Venus-like \ce{CO2} abundance is required to achieve a strong detection at SNR$=14$. In Figure~\ref{fig:heatmaps}c, at 1.52$\mu$m, Venus-like, Archean-like, and Proterozoic-like \ce{CO2} abundances achieve strong detections starting at SNRs of 8, 9, and 15, respectively. In Figure~\ref{fig:heatmaps}d, at 1.63$\mu$m, Venus-like, Archean-like, and Proterozoic-like \ce{CO2} abundances achieve strong detections at SNRs of 6, 7, and 14, respectively. In Figures~\ref{fig:heatmaps}c and d, we notice a descending staircase pattern where, as we look to the right at increasing \ce{CO2} abundances, the SNR required to achieve a strong detection is lower than the last. Furthermore, as seen in Figures~\ref{fig:heatmaps}a and b at 1.21$\mu$m and 1.44$\mu$m, strong detections are only achievable at a Venus-like \ce{CO2} abundance compared to Figures~\ref{fig:heatmaps}c and d at 1.52$\mu$m and 1.63$\mu$m where Proterozoic-like and Archean-like \ce{CO2} abundances are also able to achieve strong detections. Therefore, we confirm that longer wavelengths are necessary to achieve strong \ce{CO2} detections for a wider range of \ce{CO2} abundances - as low as 1$\times10^{-2}$ VMR.

As we consider the optimal long-wavelength cut-off for the HWO coronagraph, we want to be able to detect and characterize a wide range of \ce{CO2} abundances that existed throughout Earth's history - not just a planetary archetype similar to Venus. Only in Figures~\ref{fig:heatmaps}c and d at 1.52$\mu$m and 1.63$\mu$m are we able to achieve strong \ce{CO2} detections with \ce{CO2} abundances lower than a Venus-like abundance. Thus, moving forward, it is essential to analyze the wavelength space between 1.52-1.63$\mu$m in greater detail to ensure that we can study a variety of diverse worlds.

However, Figure~\ref{fig:heatmaps}a and b provide valuable insight on another essential topic of discussion - overlapping spectral features. Generally, we assume that deeper spectral features for a molecular species will naturally result in lower SNR requirements for detection - however, we see the opposite happening in Figure~\ref{fig:heatmaps}a and b. Looking at the left panel in Figure~\ref{fig:co} where there are only spectral features coming from \ce{CO2}, there is a deeper \ce{CO2} feature at 1.44$\mu$m compared to 1.21$\mu$m. Despite this, we see in Figure~\ref{fig:heatmaps}a at 1.21$\mu$m, a SNR$=$9 is required to achieve a strong \ce{CO2} detection compared to Figure~\ref{fig:heatmaps}b at 1.44$\mu$m, where a SNR$=$14 is required to achieve a strong \ce{CO2} detection with a Venus-like \ce{CO2} abundance. Thus, we see that a more shallow feature achieves a strong detection at a lower SNR; therefore, there must be an outside factor negatively impacting \ce{CO2} detectability at 1.44$\mu$m. As we zoom out and compare this to a modern-Earth spectrum where all molecular features are incorporated with a Venus-like \ce{CO2} abundance in Figure~\ref{fig:intro}, we see that at 1.44$\mu$ there is a massive \ce{H2O} feature overlapping with \ce{CO2}, compared to 1.21$\mu$ where \ce{CO2} overlaps only with the tail end of a smaller \ce{H2O} feature. This introduces the complexity of overlapping spectral features and how this effect impacts detectability, which we subsequently investigate at a deeper level.





\begin{figure}[hbt!]
\centering
\includegraphics[scale=.75]{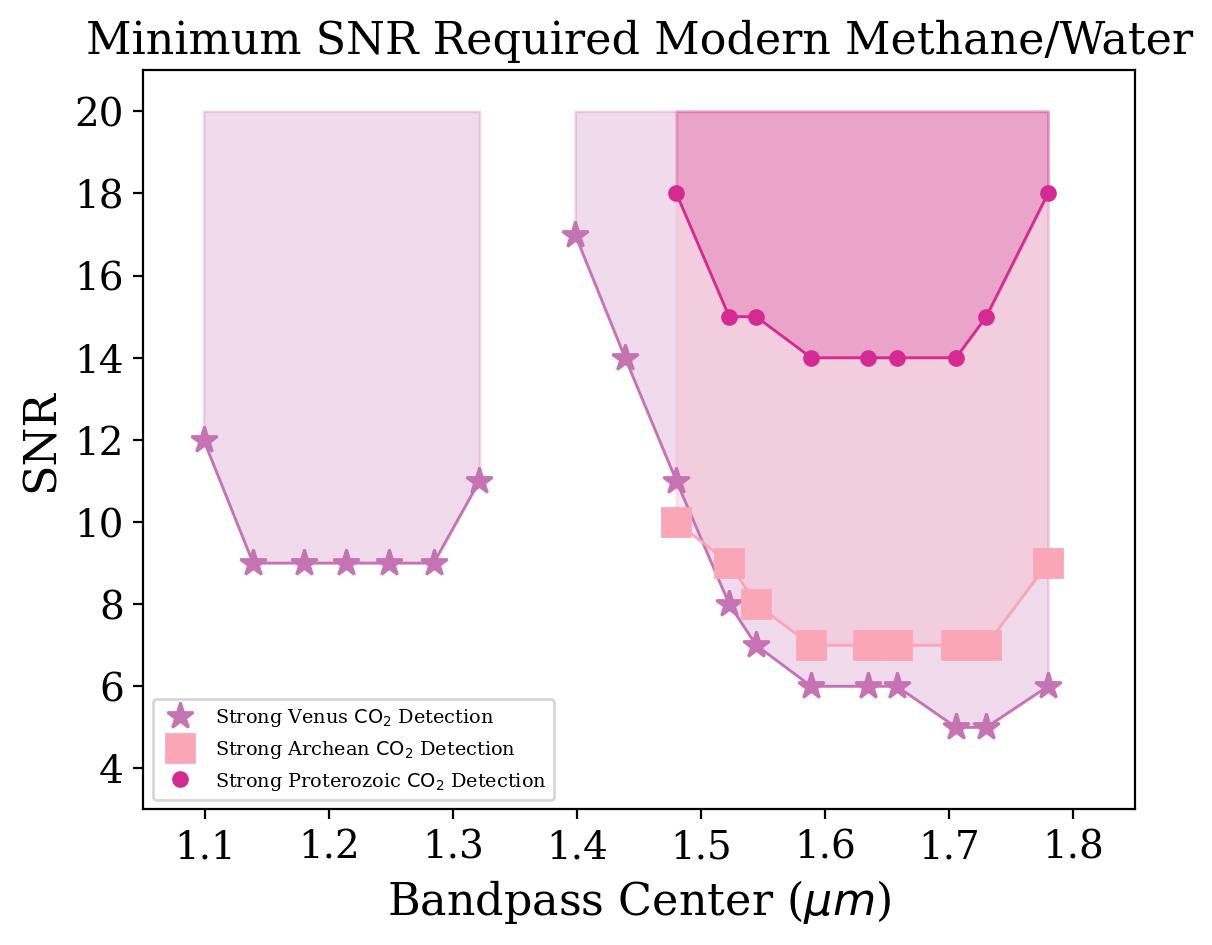}
\caption{Minimum SNR required for a strong \ce{CO2} detection at each bandpass center for the three highest \ce{CO2} abundances that achieve strong detections. Venus-like is plotted in purple stars, Archean-like is plotted in light pink squares, and Proterozoic-like is plotted in magenta circles. The shaded regions represent the spaces in which you could achieve a strong detection for each abundance in their respective colors. The x-axis is bandpass center ($\mu$m) and the y-axis is SNR.}
\label{fig:minimum_snr}
\end{figure}


Figure~\ref{fig:minimum_snr} summarizes Figure~\ref{fig:heatmaps} by showing the minimum SNR required for a strong \ce{CO2} detection at every bandpass center ($\mu$m) for the three strongly detectable \ce{CO2} abundances. There are four things we can take away from this plot: (1) only at a Venus-like \ce{CO2} abundance can we achieve a strong detection below 1.44$\mu$m, (2) no detection is possible at 1.36$\mu$m, which we can attribute to the presence of a massive \ce{H2O} feature as seen in Figure~\ref{fig:intro}, (3) the lowest and second-lowest SNRs required to achieve a strong \ce{CO2} detection occur between 1.59-1.73$\mu$m for all three \ce{CO2} abundances, and (4) at 1.78$\mu$m, we see an increase in SNR required to achieve a strong detection for all three \ce{CO2} abundances.

Our first take away indicates that if we are characterizing a planet with a Venus-like \ce{CO2} abundance with modern abundances of \ce{H2O} and \ce{CH4}, we would be able to strongly detect \ce{CO2} at most wavelengths between 1.1-1.78$\mu$m. However, to characterize a planet with lower \ce{CO2} abundances, we must look between 1.48-1.78$\mu$m. Our third and fourth points illustrate that to achieve a strong detection, the SNR requirements are lowest between 1.59-1.73$\mu$m and, in fact, increase after this point, meaning that it is unnecessary to consider going longer than a bandpass center of 1.73$\mu$m for the HWO long-wavelength cut-off. Additionally, just as we saw an irregularity at 1.44$\mu$m in Figure~\ref{fig:heatmaps}, we see another in Figure~\ref{fig:minimum_snr} at 1.36$\mu$m as stated in our second point - confirming that we need to investigate the impact of overlapping spectral features in greater detail.

\begin{figure*}
\centering
\gridline{\fig{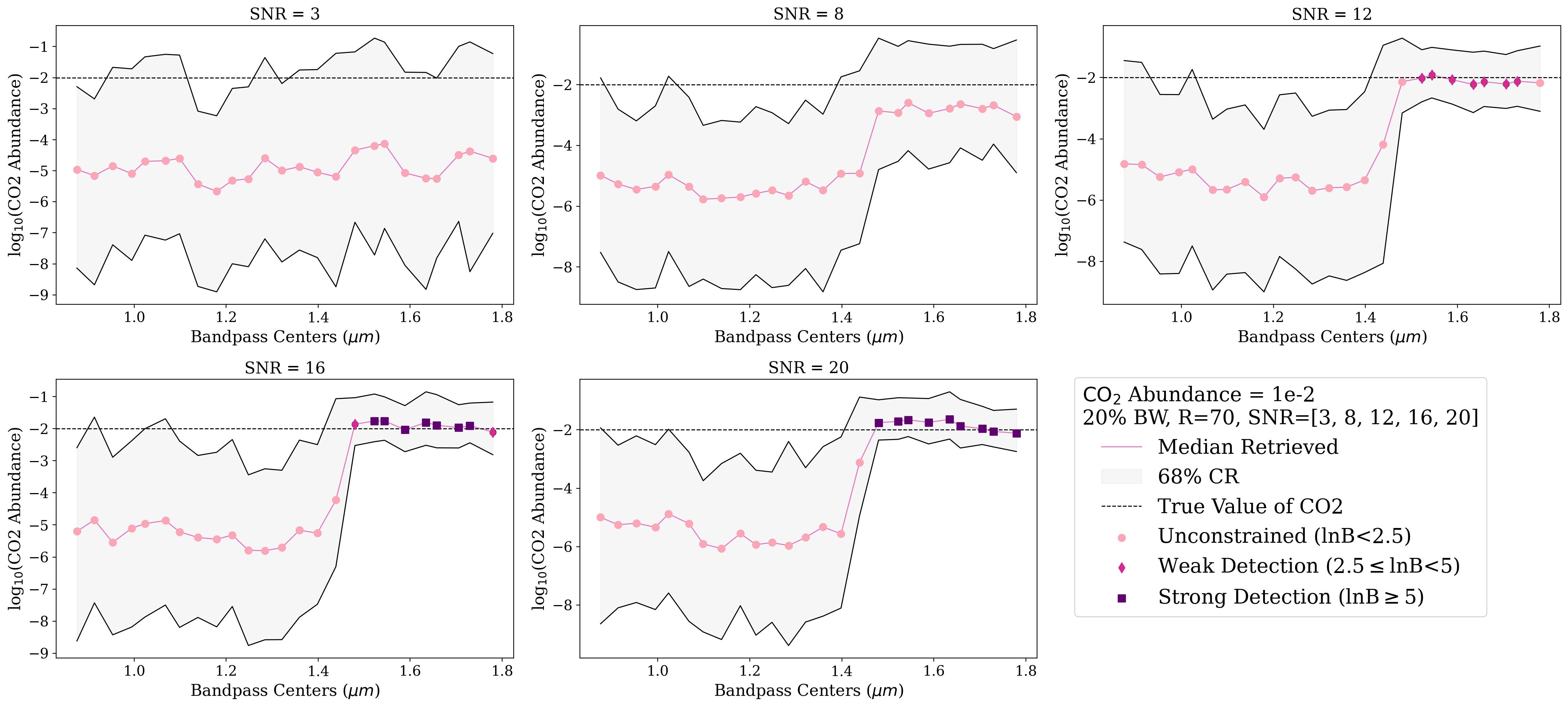}{.74\textwidth}{\vspace{-0.35cm}(a) Proterozoic level \ce{CO2} abundance (1$\times10^{-2}$ VMR)}}
\gridline{\fig{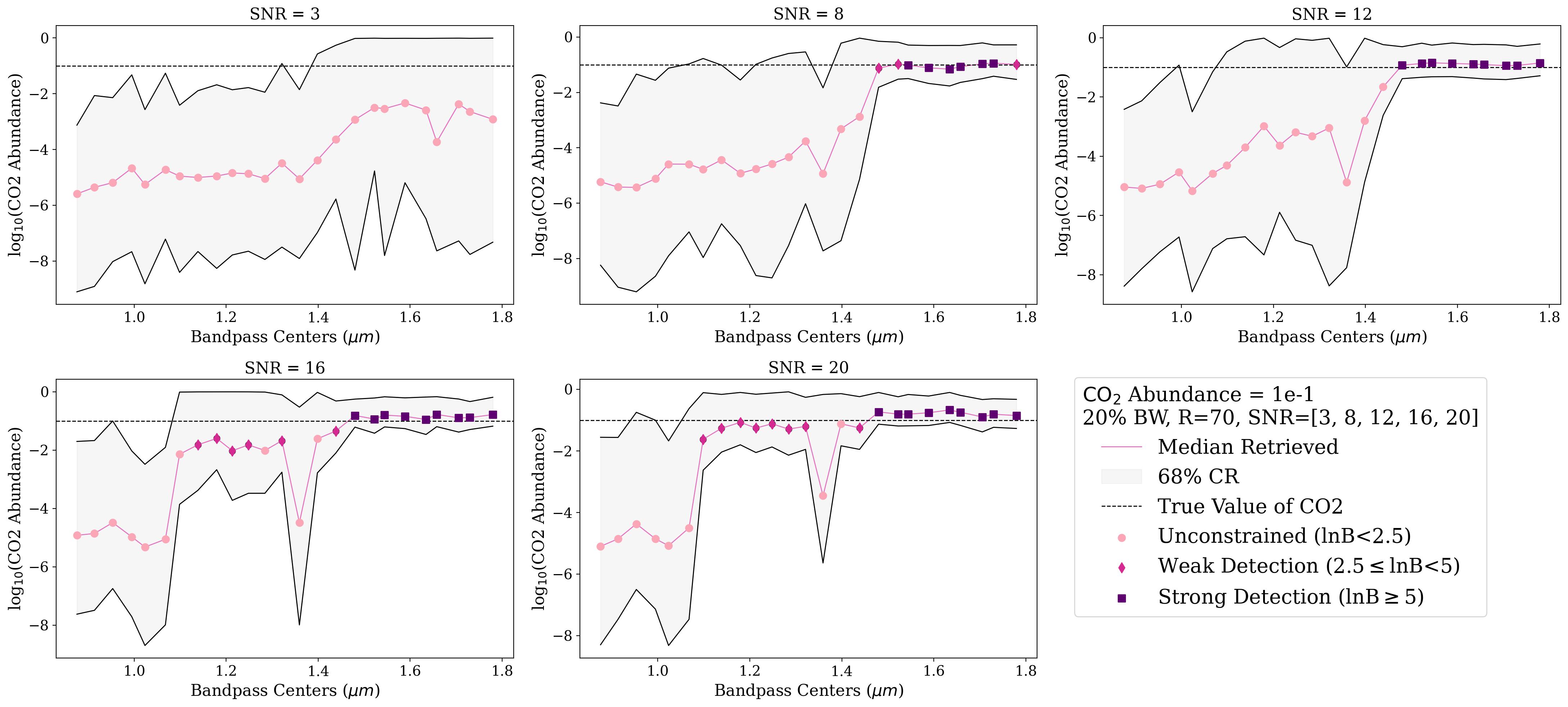}{.74\textwidth}{\vspace{-0.35cm}(b) Archean level \ce{CO2} abundance (1$\times10^{-1}$ VMR)}}
\gridline{\fig{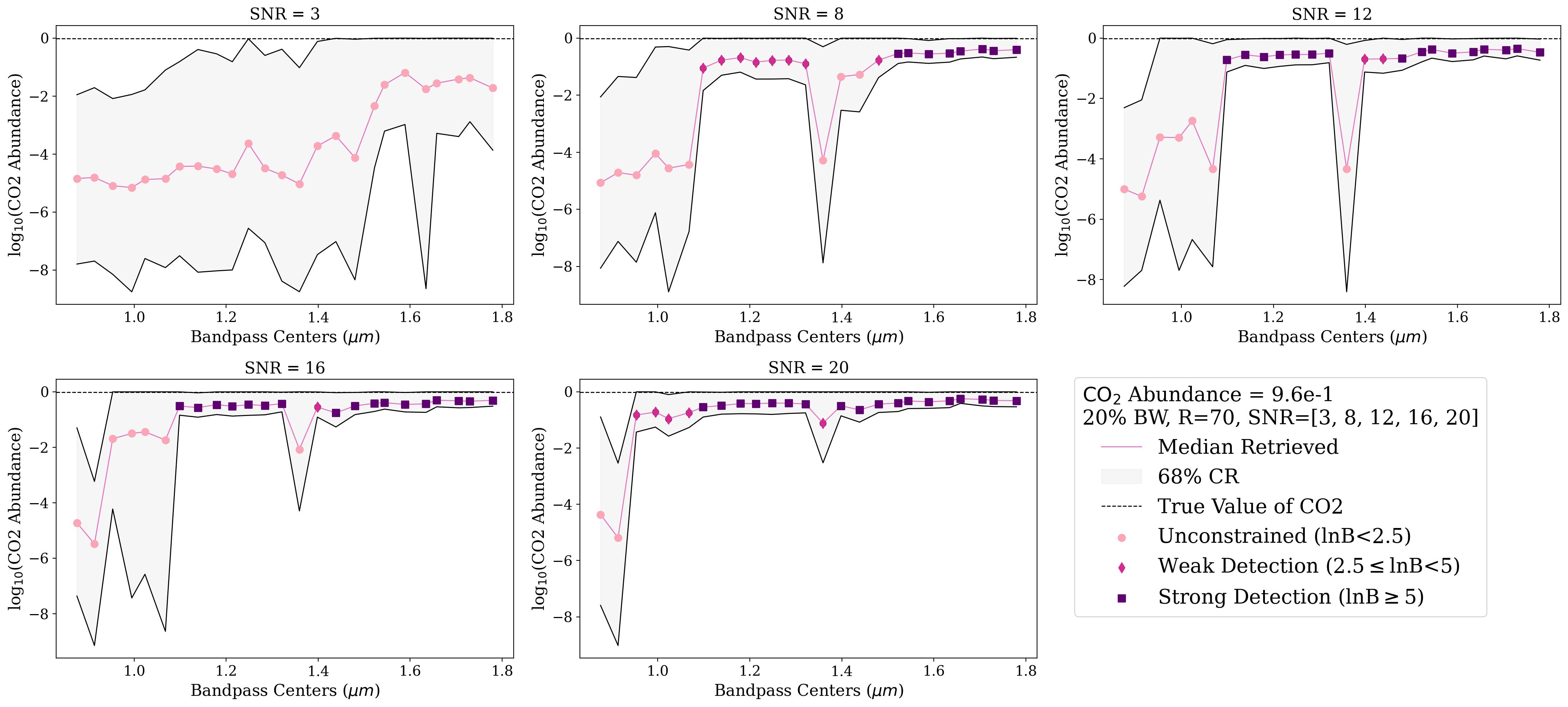}{.74\textwidth}{\vspace{-0.35cm}(c) Venus level \ce{CO2} abundance (9.6$\times10^{-1}$ VMR)}}
\caption{Detectability strength of (a) Proterozoic-like, (b) Archean-like, and (c) Venus-like \ce{CO2} abundances between 0.8-2.0$\mu$m for a subset of SNRs, in this case 6, 10, 14, 18, and 20. The light pink circles portray an unconstrained detection, the magenta diamonds portray a weak detection, and the purple squares portray a strong detection which align with the $\mathrm{lnB}$ classifications defined in Table~\ref{tab:lnB}. The 68\% credible region in shaded in grey. The true value of the \ce{CO2} abundance is marked by the horizontal dashed black line.}
\label{fig:detectability}
\end{figure*}

To allow for a more in depth analysis, we create Figure~\ref{fig:detectability} which shows the varying strength of detectability and the accuracy of the retrievals in greater detail from Proterozoic-like, Archean-like, and Venus-like \ce{CO2} abundances, respectively. Not only does Figure~\ref{fig:detectability} confirm our findings from Figure~\ref{fig:heatmaps} and Figure~\ref{fig:minimum_snr} that shorter wavelengths and lower SNRs can be used at higher \ce{CO2} abundances - it also shows us which specific wavelengths have detectability abnormalities such as a spike or drop. In Figure~\ref{fig:detectability}a, at a Proterozoic-like \ce{CO2} abundance, the detectability plots look consistent with our previous conclusion that the number of strong detections increases at higher SNRs and at longer wavelengths without unexpected changes. However, we start to see some irregularities in Figure~\ref{fig:detectability}b which examines an Archean-like abundance; we see a prominent dip in detectability strength at SNRs of 16 and 20 at 1.36$\mu$m. Similarly, in Figure~\ref{fig:detectability}c which examines a Venus-like abundance, we see this dip in detectability at 1.36$\mu$m at SNRs of 8, 12, 16 and 20. This tells us that something is affecting our ability to strongly detect \ce{CO2} at 1.36$\mu$m, just like we saw in Figure~\ref{fig:minimum_snr}. 


\subsection[NW/AM]{No \ce{H2O} \& Archean \ce{CH4} Simulations}
\label{sec:NW-AM}

Due to the irregularities in Figure~\ref{fig:heatmaps} at 1.44$\mu$m in which a Venus-like abundance requires a higher SNR to be strongly detected compared to at 1.21$\mu$m and Figure~\ref{fig:minimum_snr} and Figure~\ref{fig:detectability} at 1.36$\mu$m where we obtain a singularity of weakened \ce{CO2} detections aligned with the massive \ce{H2O} feature in the bottom panel of Figure~\ref{fig:intro}, we hypothesize that the presence of \ce{H2O} is negatively impacting the detectability of \ce{CO2} even at the highest \ce{CO2} abundance. This result led us to question which molecules and abundances could impact the detectability of \ce{CO2}. 

\begin{figure*}
\centering
\gridline{\fig{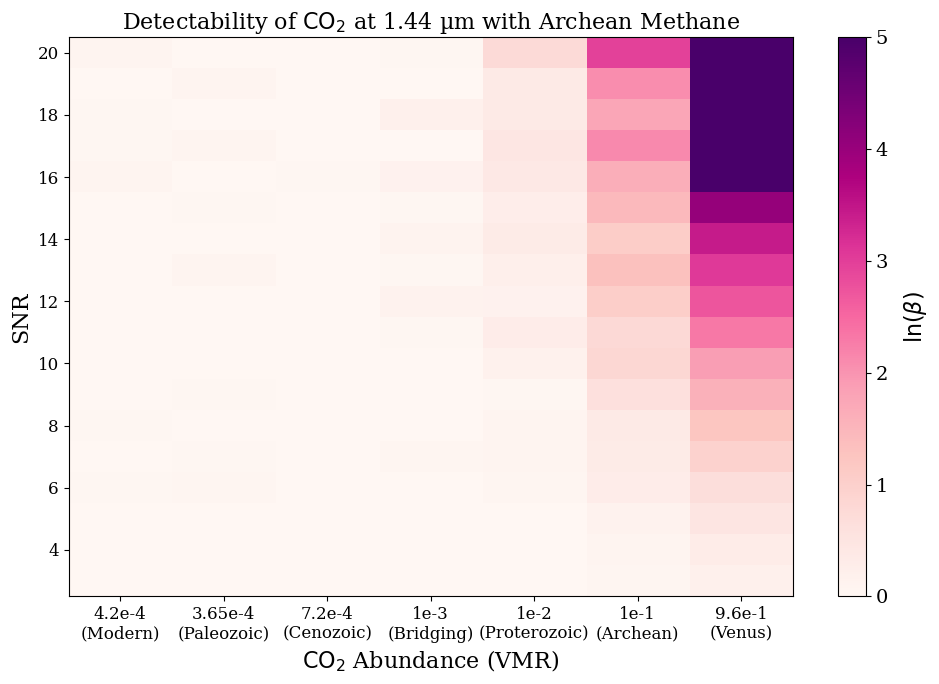}{0.33\textwidth}{\vspace{-0.35cm}(a) 1.44$\mu$m with modern \ce{H2O}/Archean \ce{CH4}}
    \fig{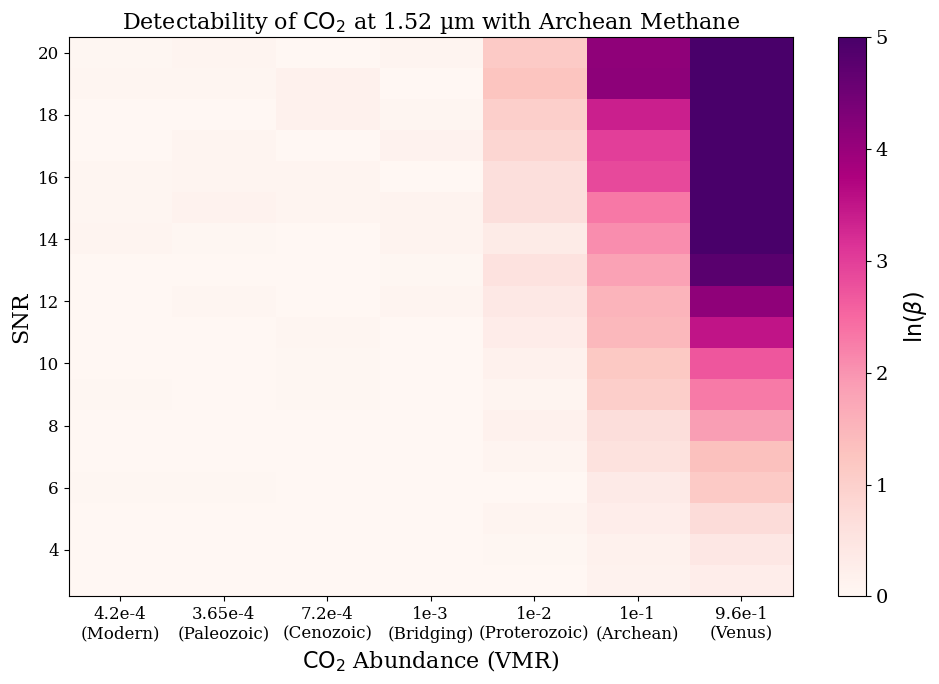}{0.33\textwidth}{\vspace{-0.35cm}(b) 1.52$\mu$m with modern \ce{H2O}/Archean \ce{CH4}}
    \fig{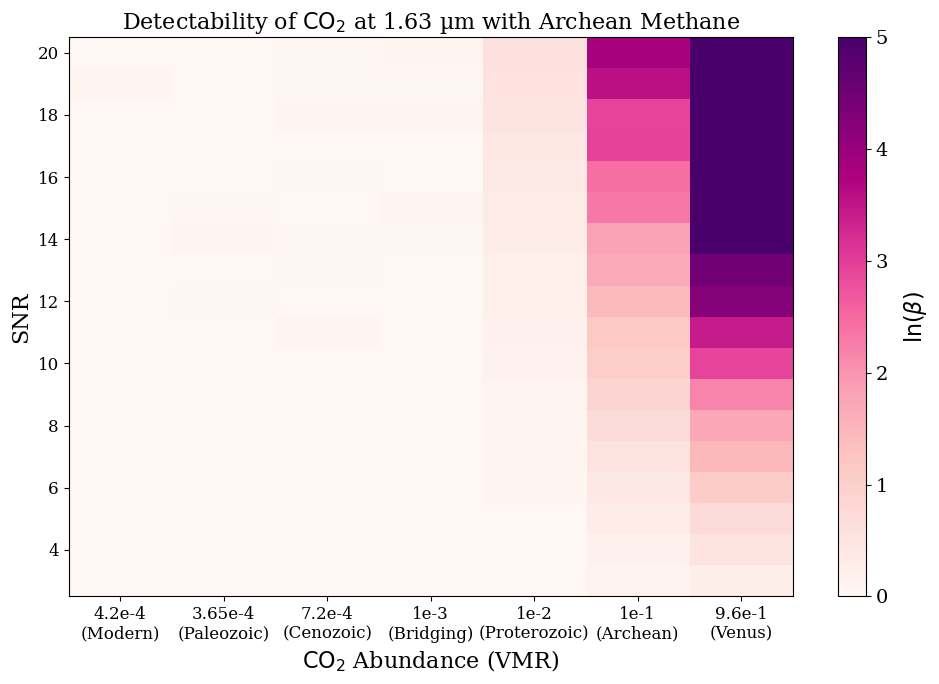}{0.33\textwidth}{\vspace{-0.35cm}(c) 1.63$\mu$m with modern \ce{H2O}/Archean \ce{CH4}}}
\gridline{\fig{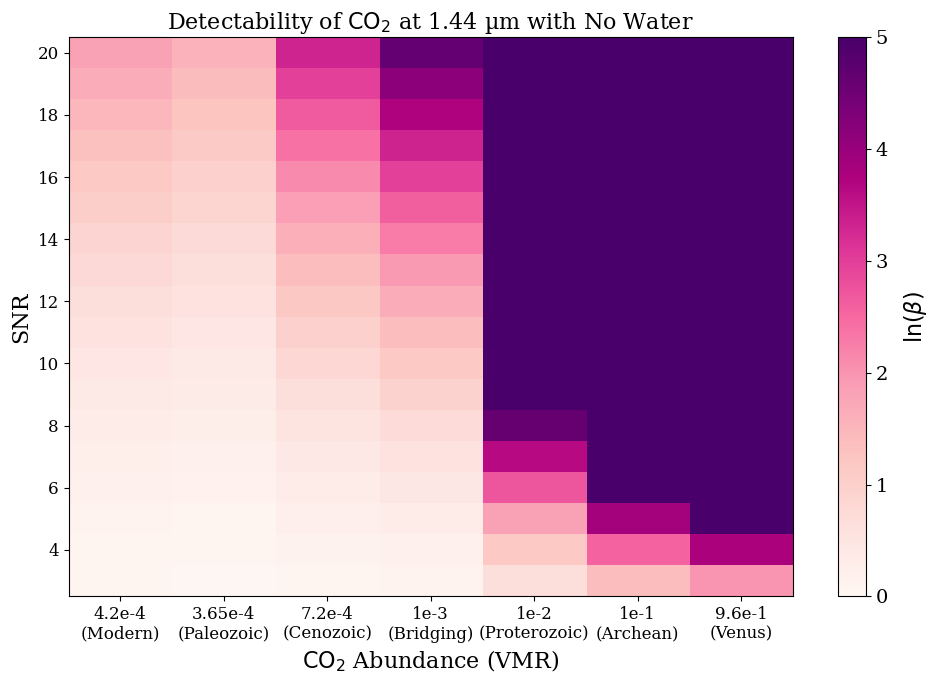}{0.33\textwidth}{\vspace{-0.35cm}(d) 1.44$\mu$m with no \ce{H2O}/modern \ce{CH4}}
    \fig{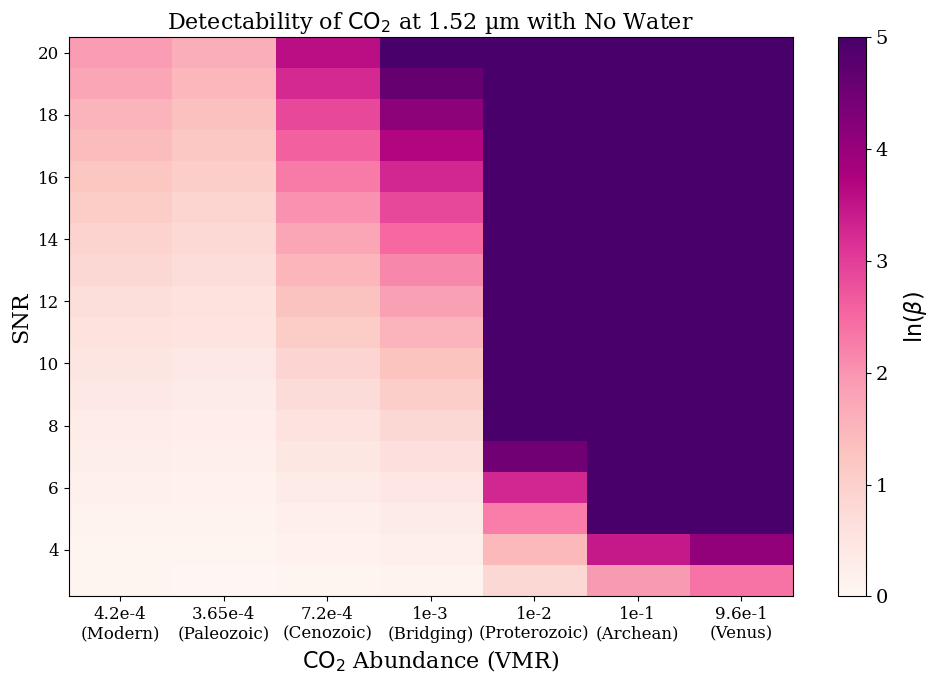}{0.33\textwidth}{\vspace{-0.35cm}(e) 1.52$\mu$m with no \ce{H2O}/modern \ce{CH4}}
    \fig{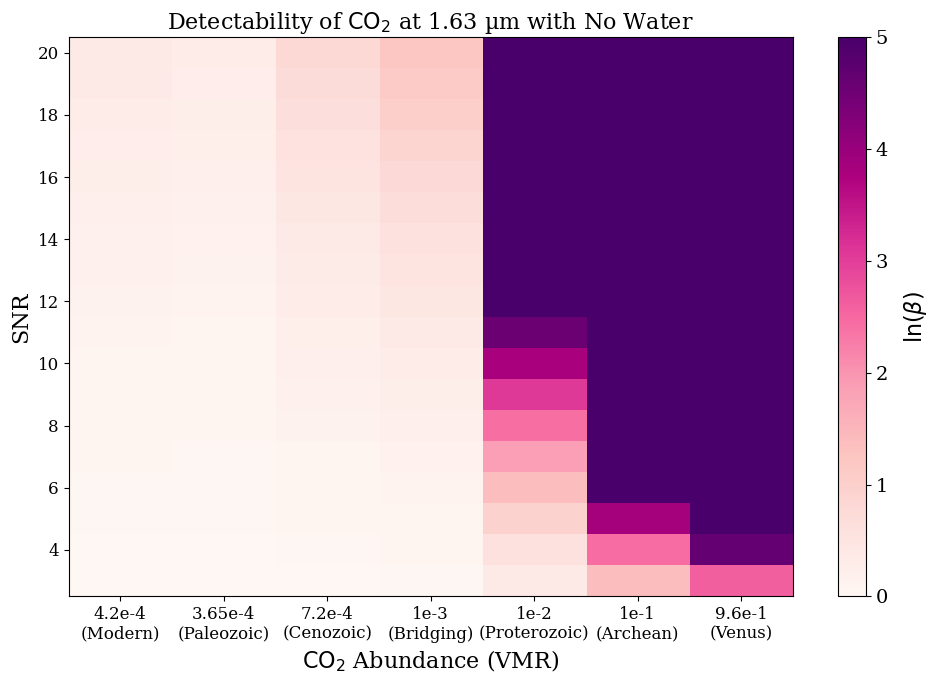}{0.33\textwidth}{\vspace{-0.35cm}(f) 1.63$\mu$m with no \ce{H2O}/modern \ce{CH4}}}
\caption{Varying detection strength of \ce{CO2} with respect to \ce{CO2} abundance (on the x-axis) and SNR (on the y-axis) at bandpass centers of 1.44$\mu$m (a \& d), 1.52$\mu$m (b \& e), and 1.63$\mu$m (c \& f). For each abundance and SNR, the color corresponds to the strength of the \ce{CO2} detection; the darker the purple, the stronger the detection, and the lighter the pink, the more unconstrained the detection. The top plots (a, b, and c) show \ce{CO2} detection strength for modern \ce{H2O} and Archean \ce{CH4}. The bottom plots (d, e, and f) show \ce{CO2} detection strength for no \ce{H2O} and modern \ce{CH4}.}
\label{fig:1.78_heatmaps}
\end{figure*}

We simulated two additional tests that investigate how replacing (1) modern \ce{CH4} with Archean \ce{CH4} (7.07$\times10^{-3}$), thereby increasing the \ce{CH4} abundance, and (2) modern \ce{H2O} with no \ce{H2O}, thereby decreasing the \ce{H2O} abundance, would change the detectability of \ce{CO2}. Thus, Figure~\ref{fig:1.78_heatmaps} allows for the analysis of the interconnected \ce{CO2}, \ce{H2O}, and \ce{CH4} spectral features and the impact they have on the detectability of \ce{CO2}.



Figure~\ref{fig:1.78_heatmaps}a at 1.44$\mu$m, Figure~\ref{fig:1.78_heatmaps}b at 1.52$\mu$m, and Figure~\ref{fig:1.78_heatmaps}c at 1.63$\mu$m with modern \ce{H2O} and Archean \ce{CH4} show that with an increased abundance of \ce{CH4}, the detectability of \ce{CO2} decreases greatly across all abundances that had strong detections with modern \ce{CH4}. For example, the minimum SNR for a strong detection increases from 14, 8, and 6 to 16, 14, and 14 for a Venus-like \ce{CO2} abundance at 1.44$\mu$m, 1.52$\mu$m, and 1.63$\mu$m, respectively. The opposite effect occurs when we set \ce{H2O} to zero in Figure~\ref{fig:1.78_heatmaps}d at 1.44$\mu$m, Figure~\ref{fig:1.78_heatmaps}e at 1.52$\mu$m, and Figure~\ref{fig:1.78_heatmaps}f at 1.63$\mu$m. Compared to the original modern \ce{H2O}/\ce{CH4} heat maps in Figure~\ref{fig:heatmaps}b (1.44$\mu$m), c (1.52$\mu$m), and d (1.63$\mu$m), there is a dramatic increase in \ce{CO2} detectability across the \ce{CO2} abundances. For example, the minimum \ce{CO2} abundance that achieves a strong \ce{CO2} detection goes from \ce{CO2} abundances of 9.6$\times10^{-1}$ VMR, 1$\times10^{-2}$ VMR, and 1$\times10^{-2}$ VMR to 1$\times10^{-2}$ VMR, 1$\times10^{-3}$ VMR, and 1$\times10^{-2}$ VMR for 1.44$\mu$m, 1.52$\mu$m, and 1.63$\mu$m, respectively.

Due to the drastic impact that the presence of \ce{H2O} and \ce{CH4} have on the detectability of \ce{CO2}, we simulated tests that vary the abundances of two different molecules at the same time. The simulations were done at 1.44$\mu$m, 1.52$\mu$m, and 1.63$\mu$m, due to the strong overlapping features of \ce{CO2} with \ce{CH4}, \ce{H2O}, or both as indicated in the bottom panel of Figure~\ref{fig:intro}. We use the abundances of \ce{CH4} and \ce{H2O} from BARBIE3 which can be found in Table~\ref{tab:all_abundances}. It is important to note that we only ran \ce{CO2}-\ce{H2O} and \ce{CO2}-\ce{CH4} tests because BARBIE3 analyzes the relationship between \ce{H2O} and \ce{CH4}. Thus, BARBIE3 and this together give a comprehensive study of \ce{CO2}, \ce{CH4}, and \ce{H2O} and the relationships between the three molecules in the NIR as functions of abundance, wavelength, and SNR.


\subsection[CO2/CH4/H2O]{\ce{CO2}-\ce{CH4} \& \ce{CO2}-\ce{H2O} Simulations}
\label{sec:dual-tests}


\begin{figure*}
\centering
\gridline{\fig{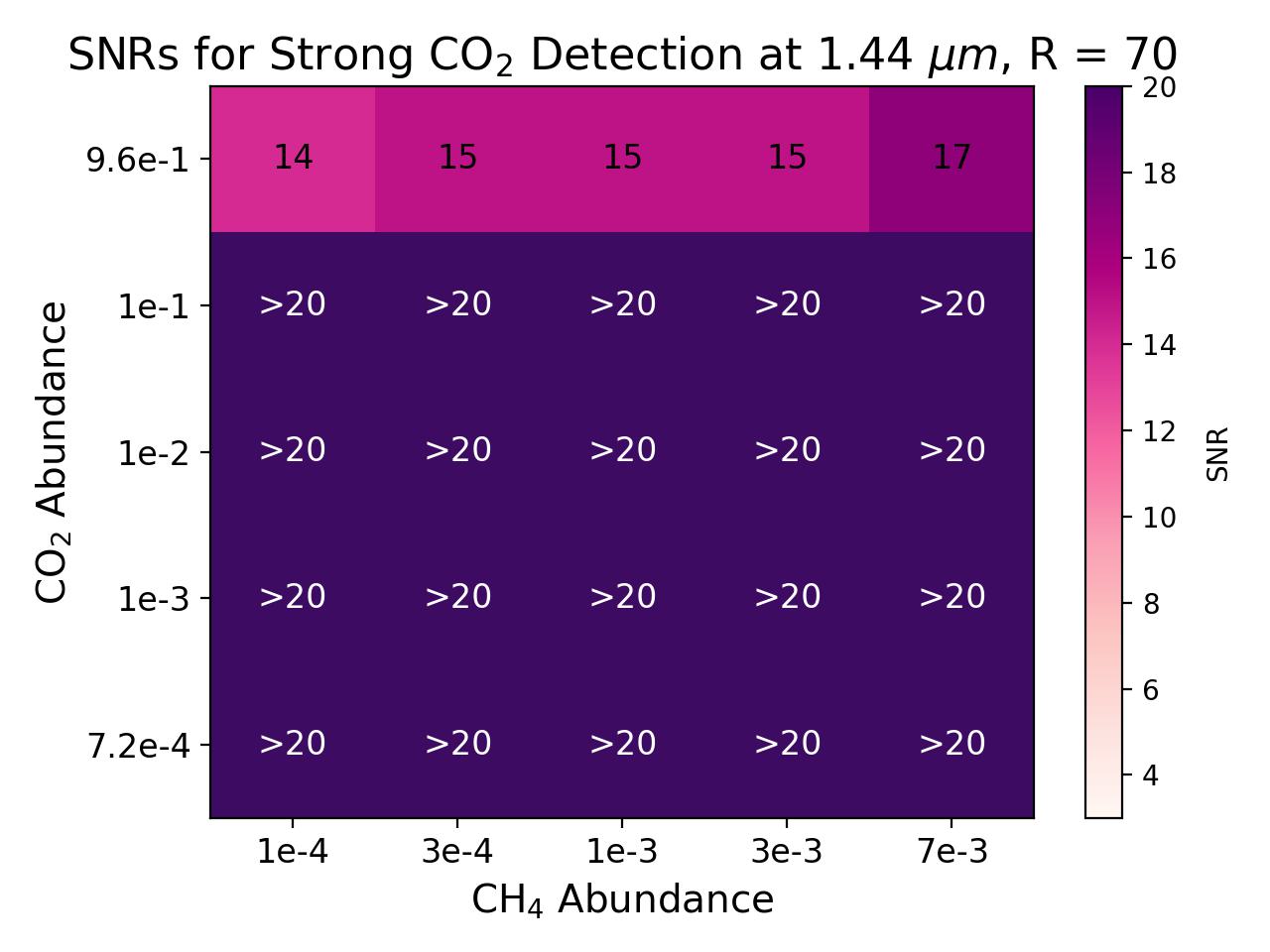}{0.33\textwidth}{\vspace{-0.2cm}(a) Detectability of \ce{CO2} at 1.44$\mu$m}
    \fig{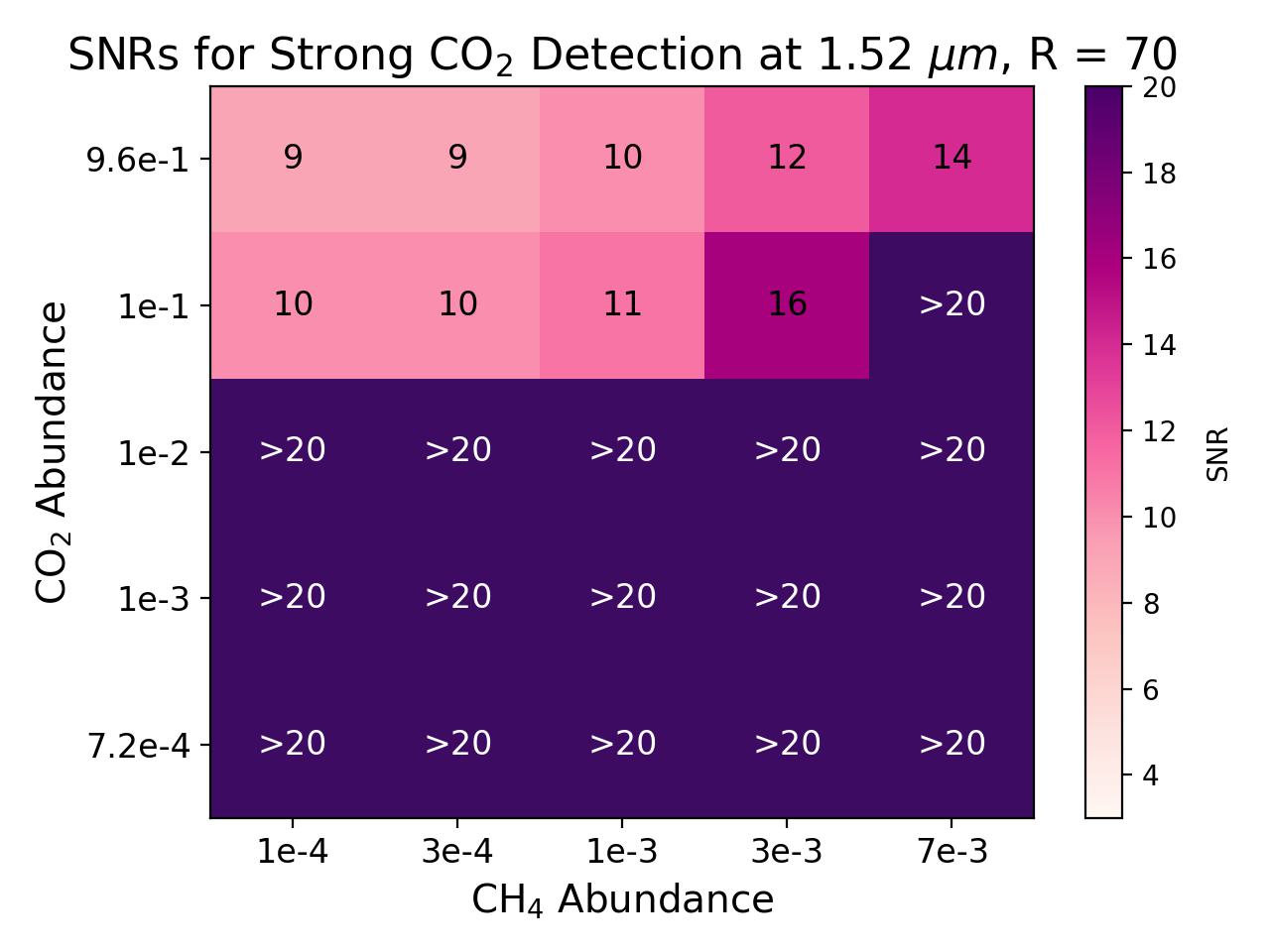}{0.33\textwidth}{\vspace{-0.2cm}(b) Detectability of \ce{CO2} at 1.52$\mu$m}
    \fig{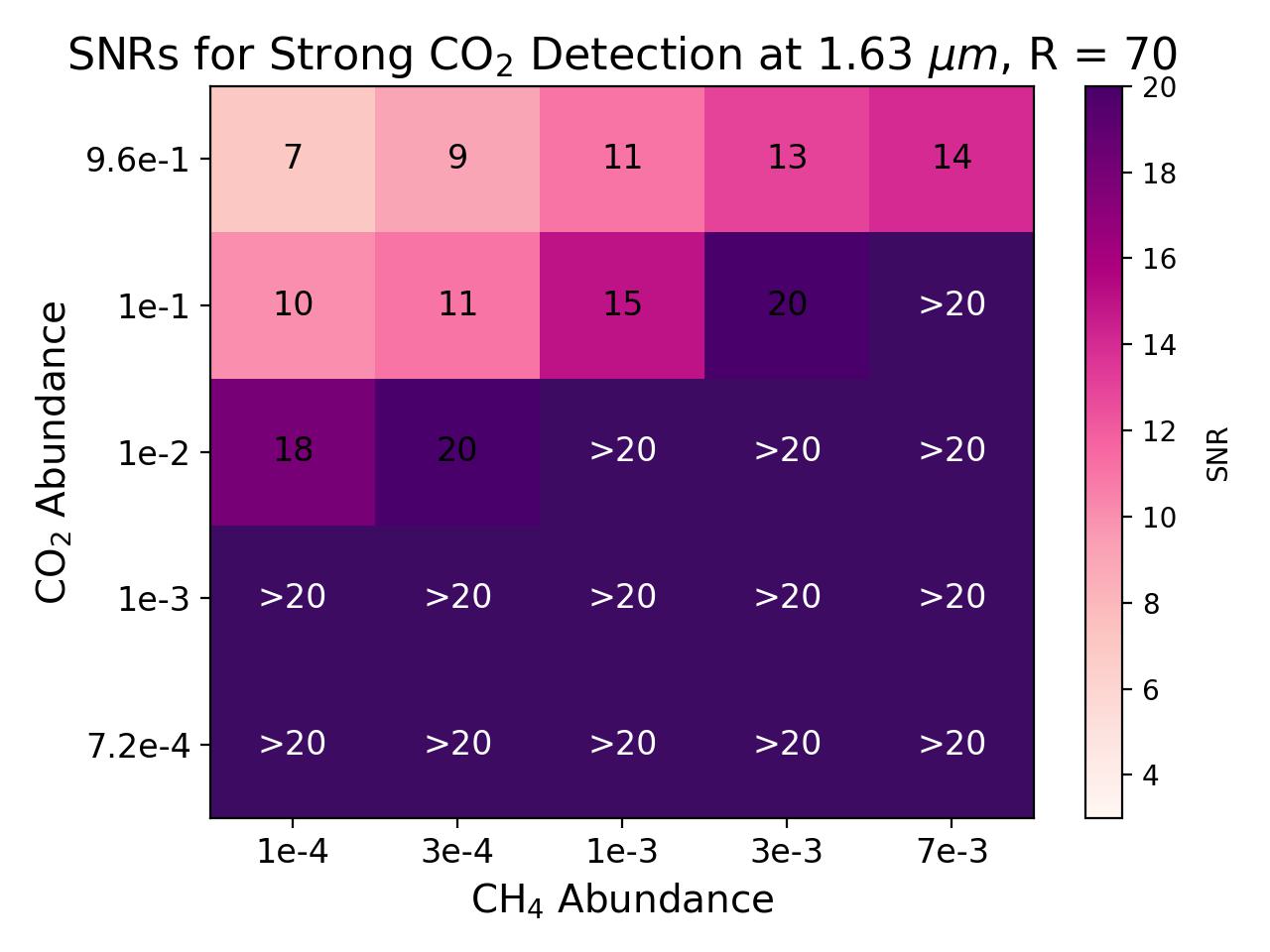}{0.33\textwidth}{\vspace{-0.2cm}(c) Detectability of \ce{CO2} at 1.63$\mu$m}}
\gridline{\fig{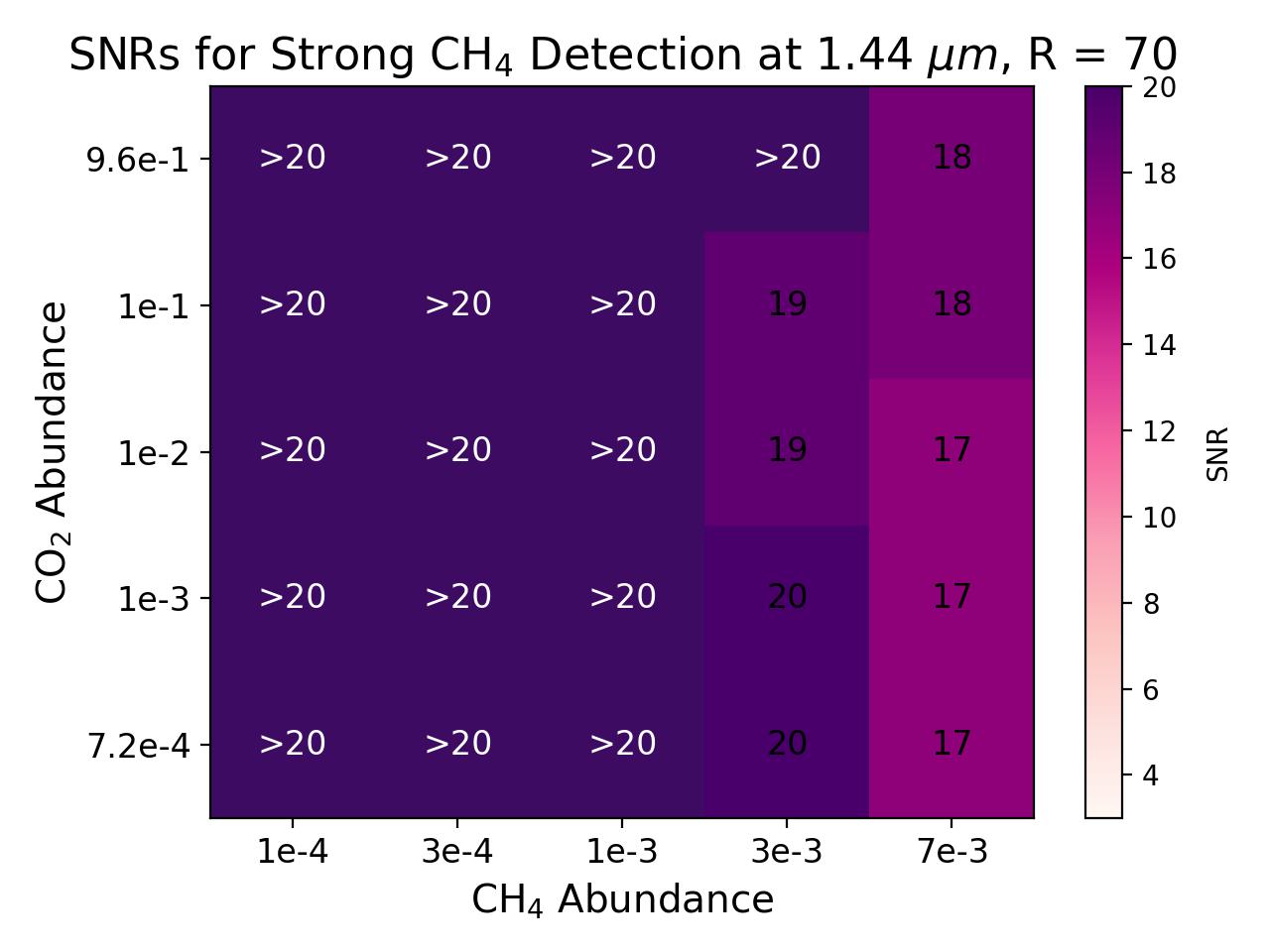}{0.33\textwidth}{\vspace{-0.2cm}(d) Detectability of \ce{CH4} at 1.44$\mu$m}
    \fig{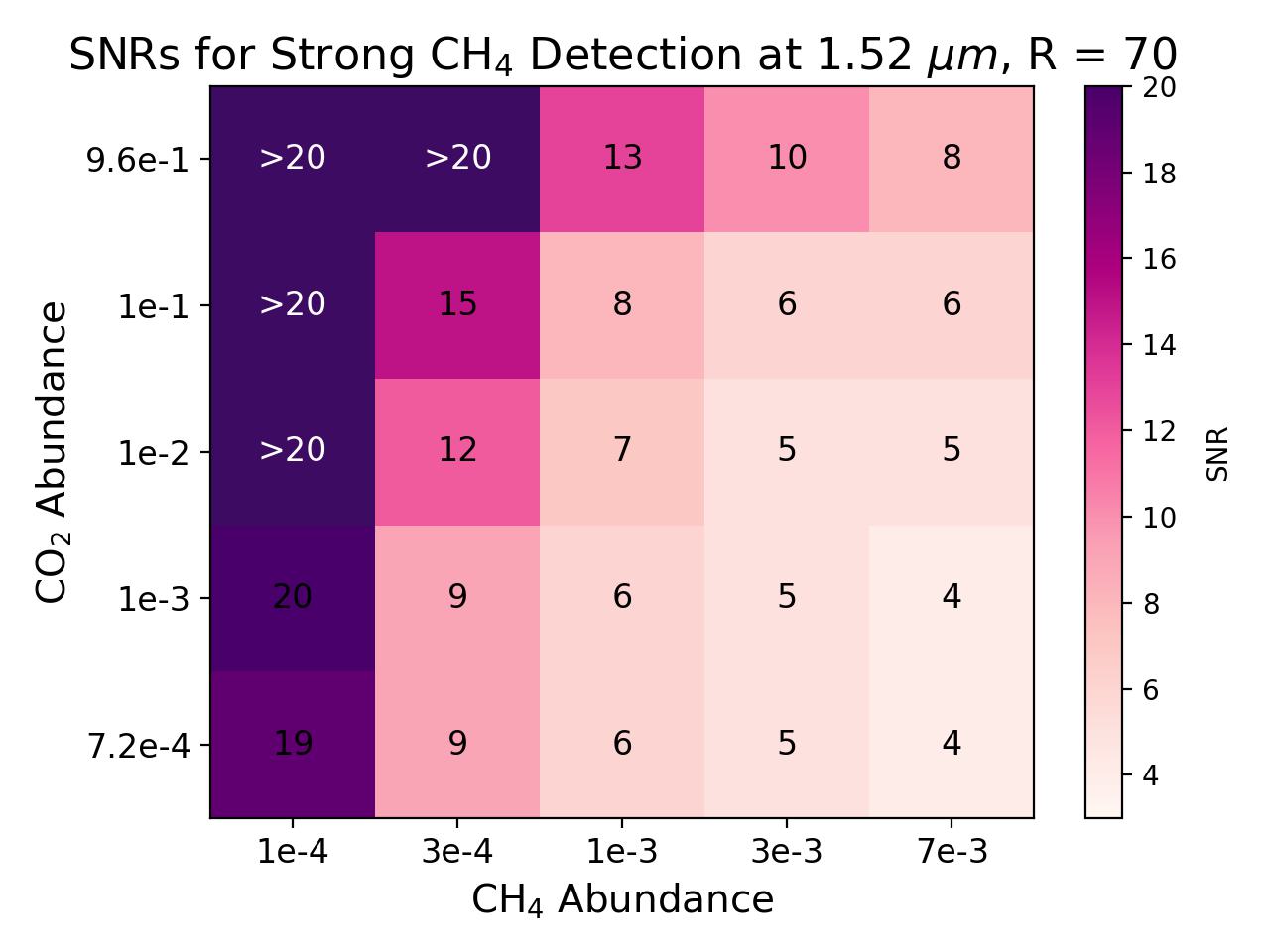}{0.33\textwidth}{\vspace{-0.2cm}(e) Detectability of \ce{CH4} at 1.52$\mu$m}
    \fig{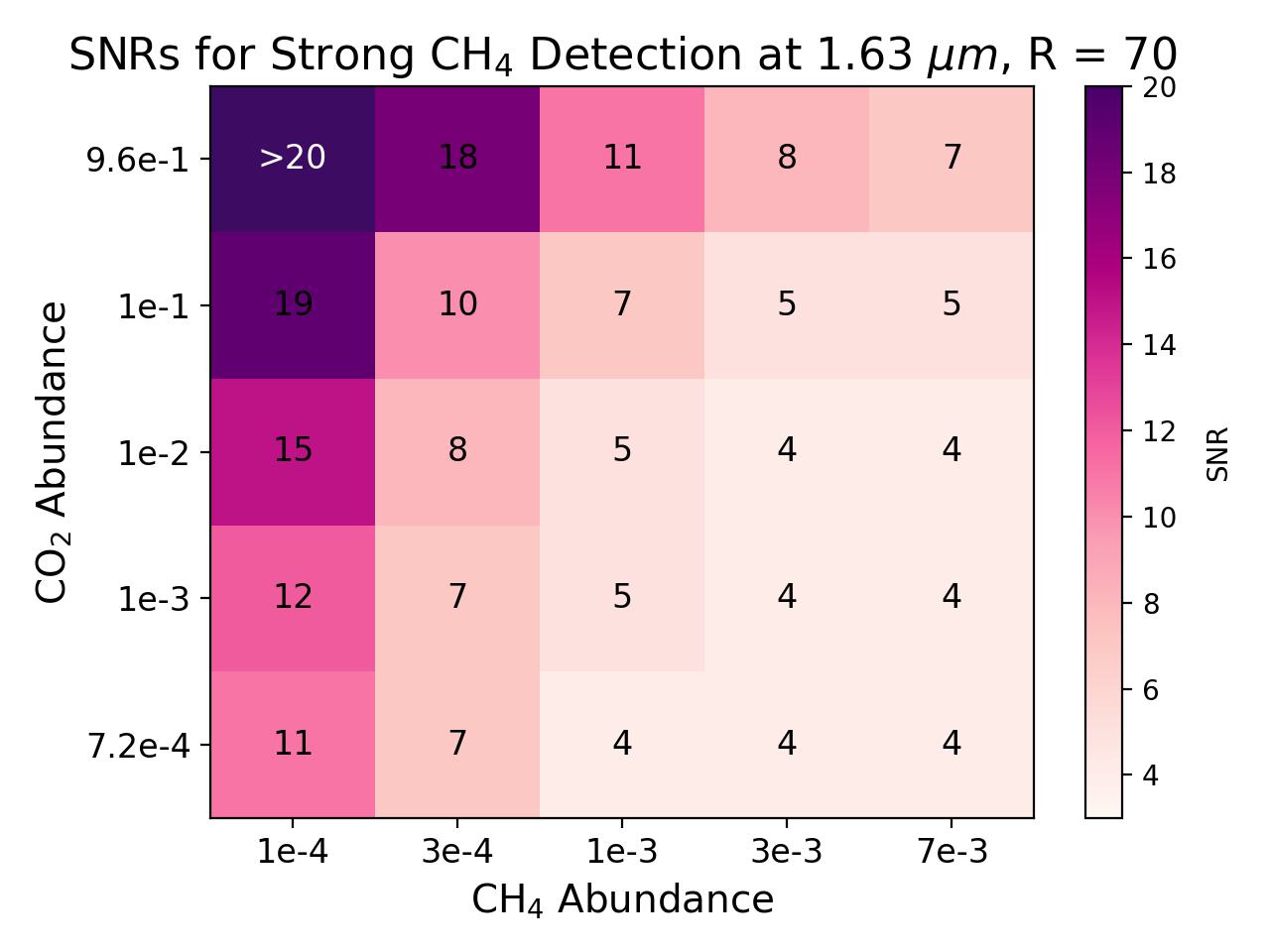}{0.33\textwidth}{\vspace{-0.2cm}(f) Detectability of \ce{CH4} at 1.63$\mu$m}}
\caption{Top: Required SNR for a strong \ce{CO2} detection at 1.44$\mu$m (a), 1.52$\mu$m (b), and 1.63$\mu$m (c). Bottom: Required SNR for a strong \ce{CH4} detection at 1.44$\mu$m (d), 1.52$\mu$m (e), and 1.63$\mu$m (f). For each \ce{CO2}-\ce{CH4} combination, the necessary SNR required for a strong detection is shown, where darker colors correlate to higher SNRs and lighter colors correlate to lower SNRs. \ce{CH4} abundances are plotted on the x-axis and \ce{CO2} abundances are plotted on the y-axis. \ce{CH4} and \ce{CO2} abundances are taken from Table~\ref{tab:all_abundances}.}
\label{fig:ch4_heatmaps}
\end{figure*}

\begin{figure*}
\centering
\gridline{\fig{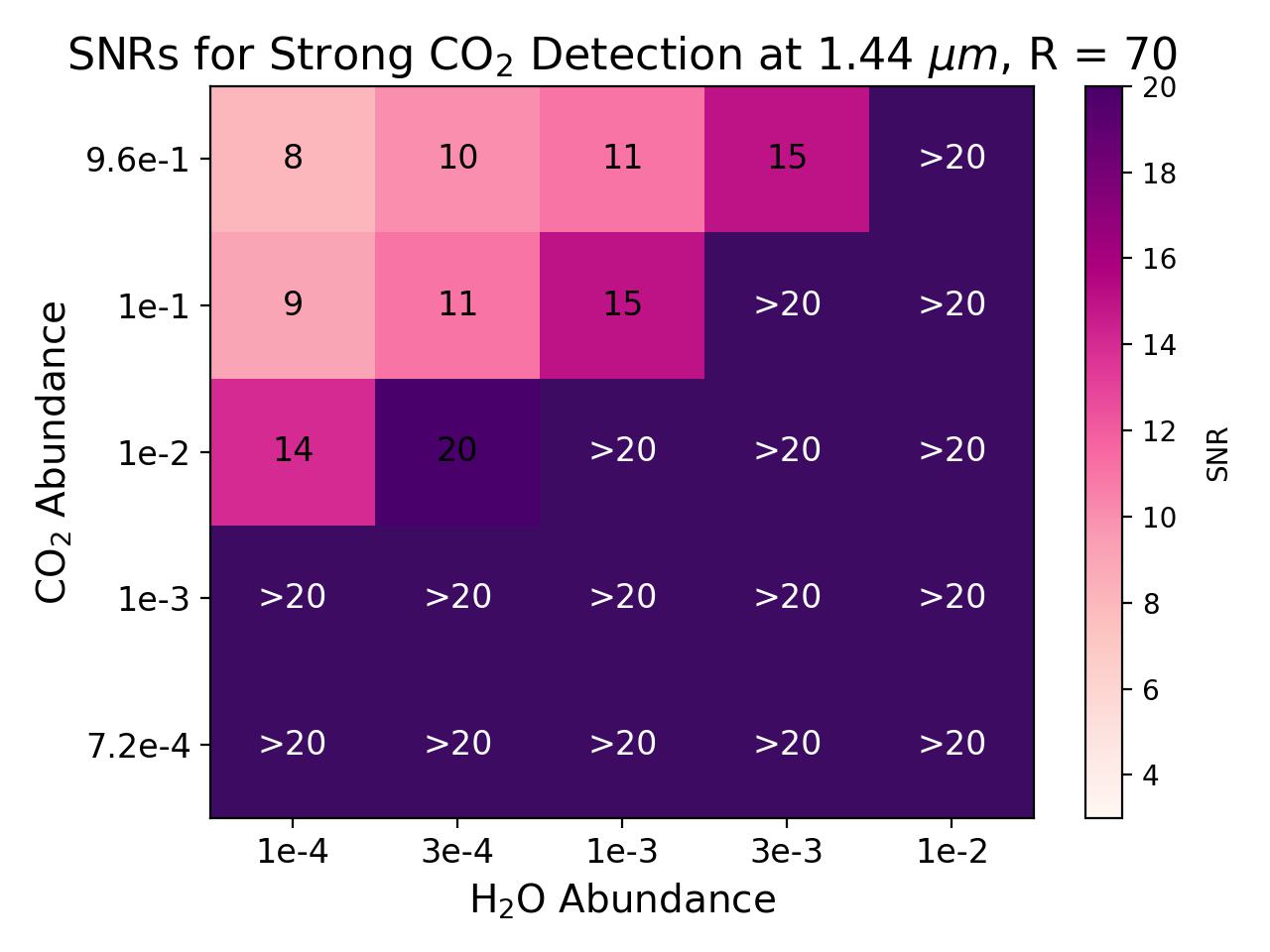}{0.33\textwidth}{\vspace{-0.2cm}(a) Detectability of \ce{CO2} at 1.44$\mu$m}
    \fig{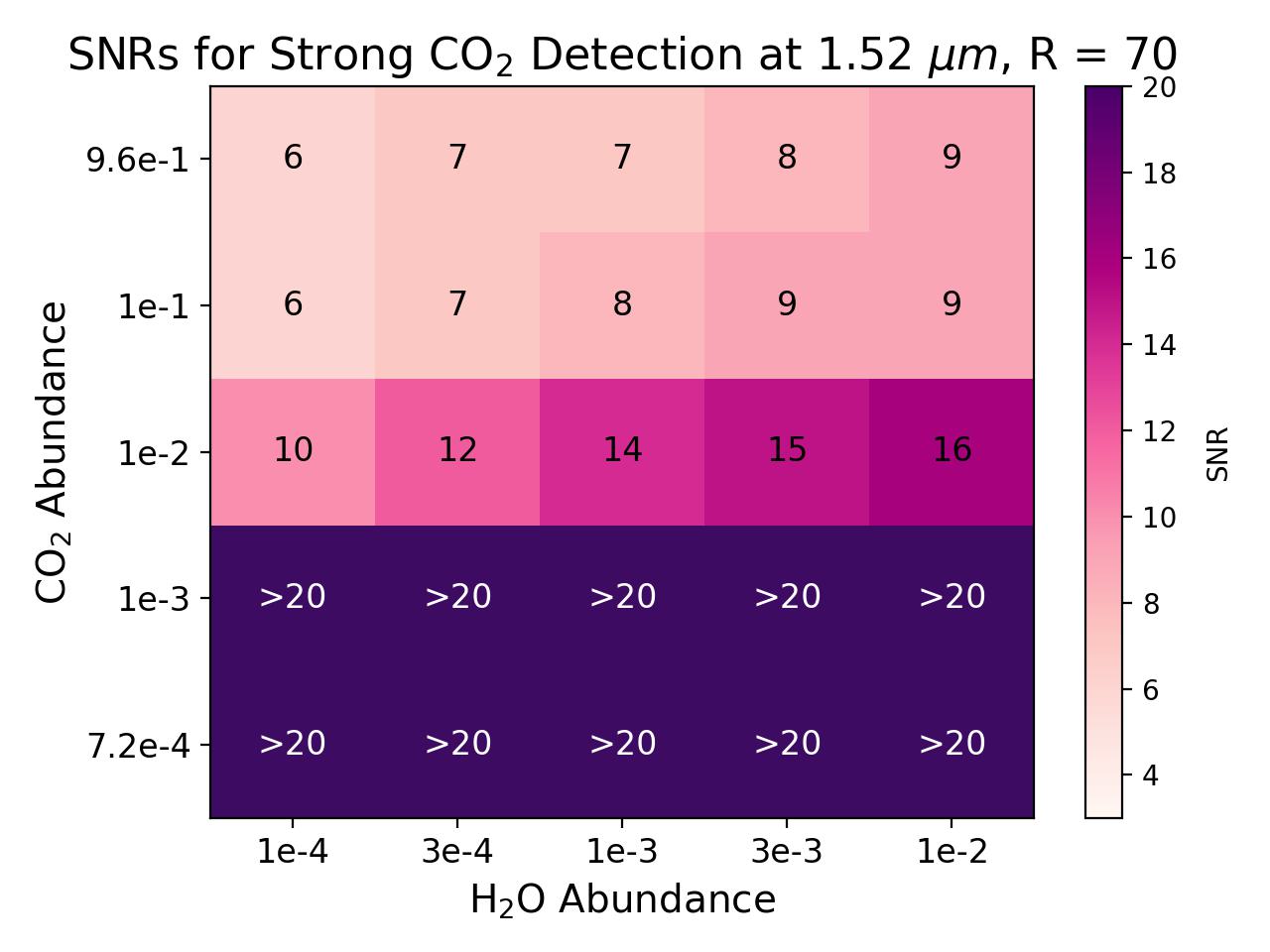}{0.33\textwidth}{\vspace{-0.2cm}(b) Detectability of \ce{CO2} at 1.52$\mu$m}
    \fig{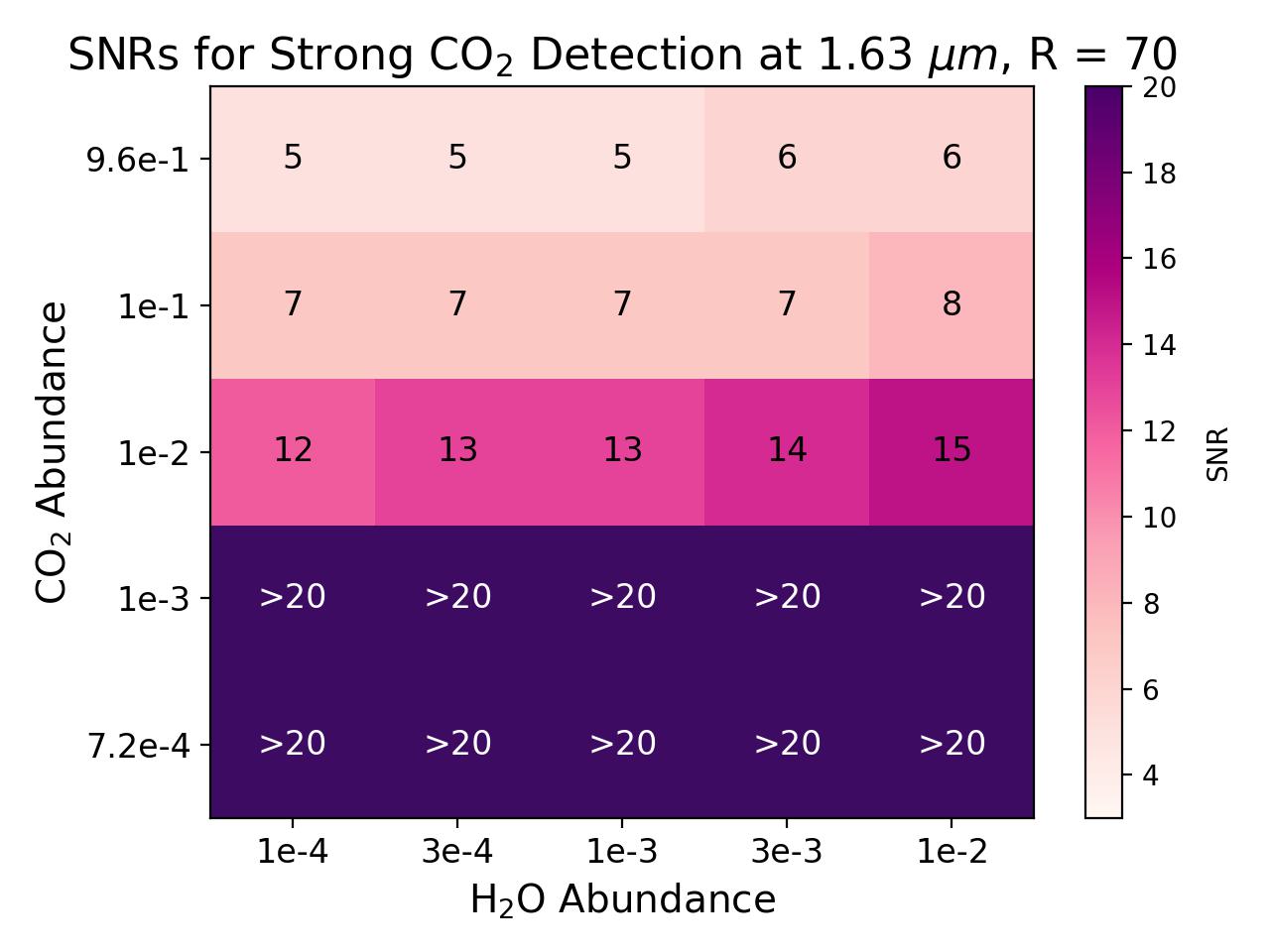}{0.33\textwidth}{\vspace{-0.2cm}(c) Detectability of \ce{CO2} at 1.63$\mu$m}}
\gridline{\fig{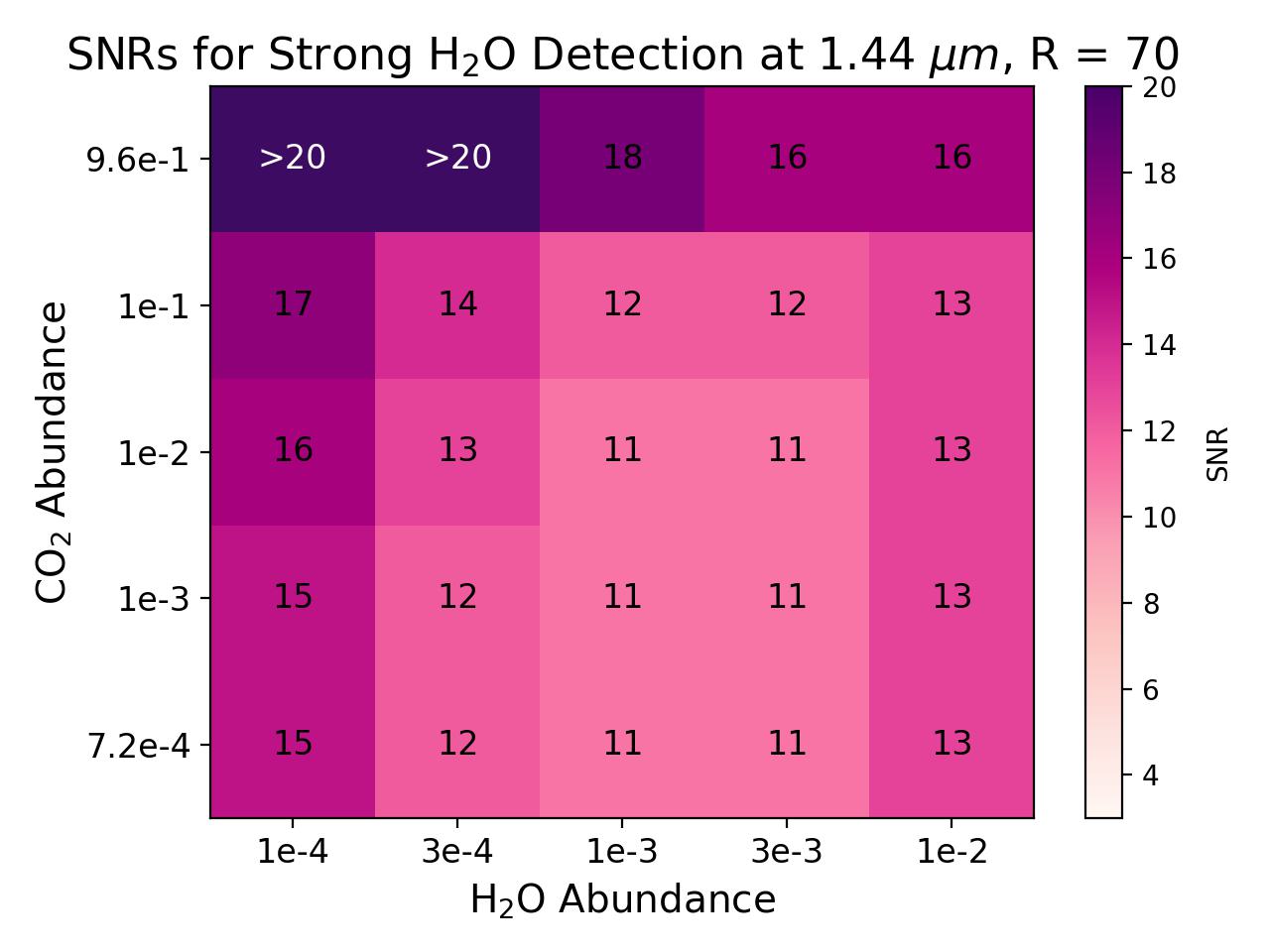}{0.33\textwidth}{\vspace{-0.2cm}(d) Detectability of \ce{H2O} at 1.44$\mu$m}
    \fig{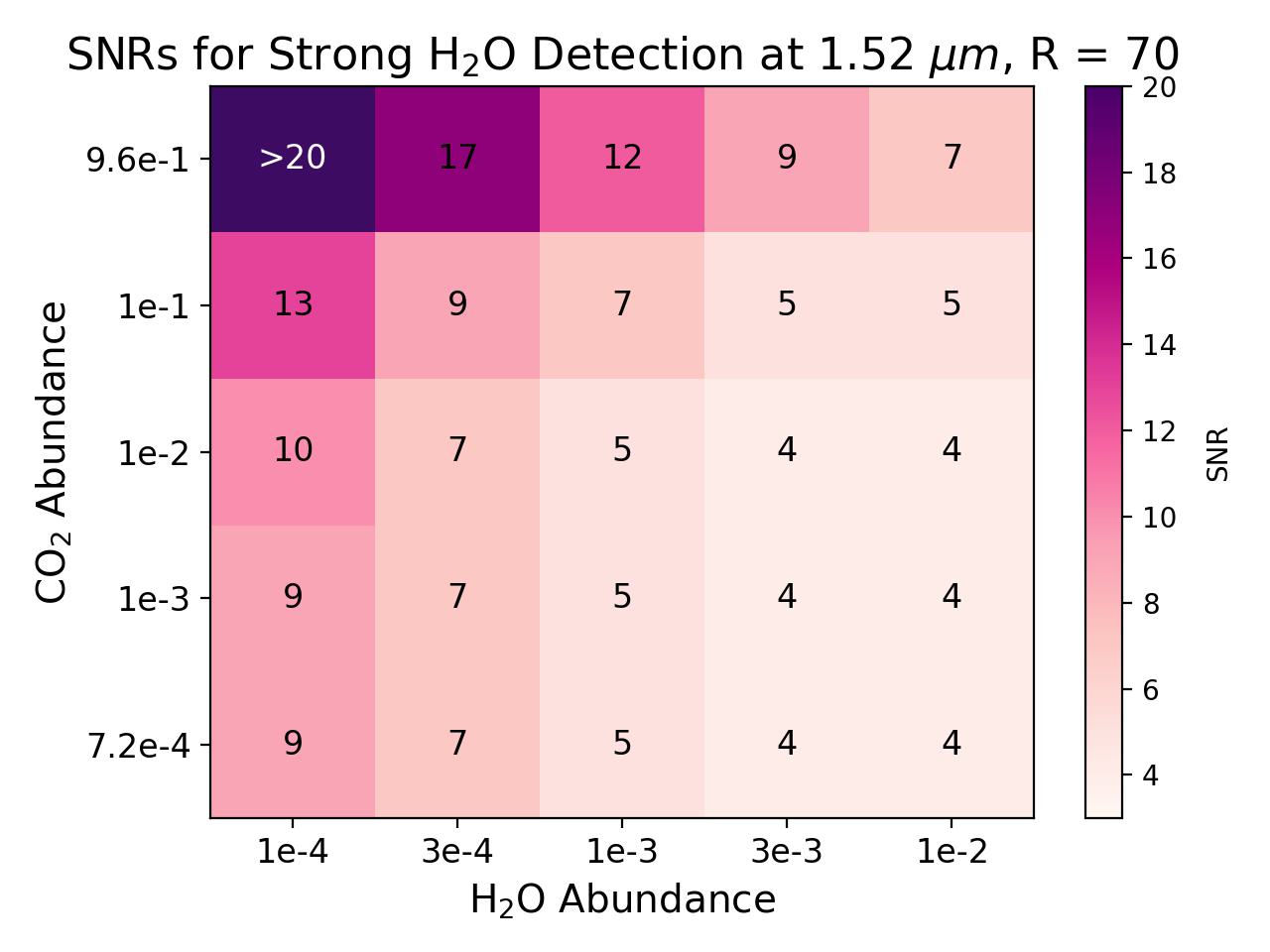}{0.33\textwidth}{\vspace{-0.2cm}(e) Detectability of \ce{H2O} at 1.52$\mu$m}
    \fig{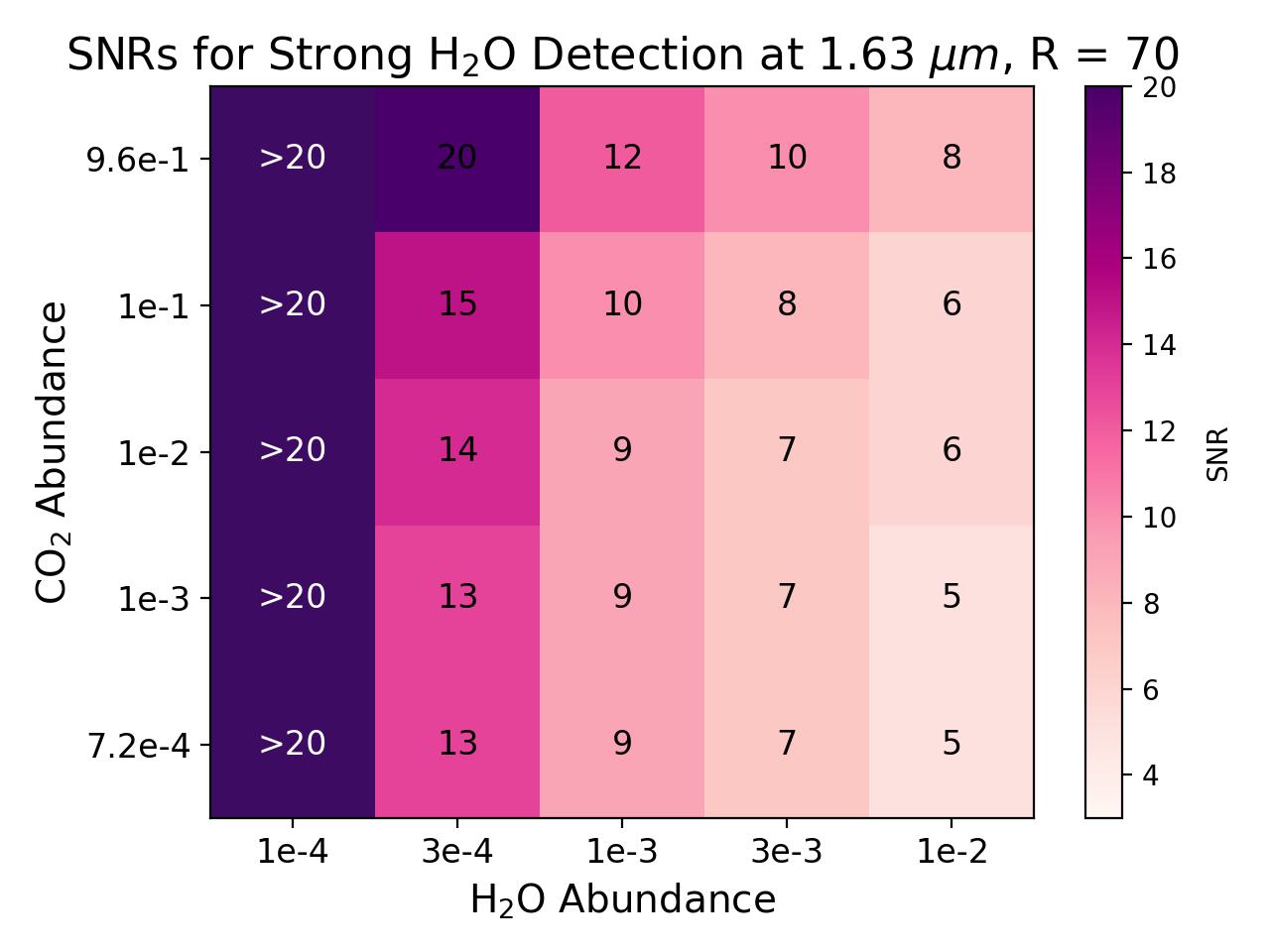}{0.33\textwidth}{\vspace{-0.2cm}(f) Detectability of \ce{H2O} at 1.63$\mu$m}}
\caption{Top: Required SNR for a strong \ce{CO2} detection at 1.44$\mu$m (a), 1.52$\mu$m (b), and 1.63$\mu$m (c). Bottom: Required SNR for a strong \ce{H2O} detection at 1.44$\mu$m (d), 1.52$\mu$m (e), and 1.63$\mu$m (f). For each \ce{CO2}-\ce{H2O} combination, the necessary SNR required for a strong detection is shown, where darker colors correlate to higher SNRs and lighter colors correlate to lower SNRs. \ce{H2O} abundances are plotted on the x-axis and \ce{CO2} abundances are plotted on the y-axis. \ce{CH4} and \ce{H2O} abundances are taken from Table~\ref{tab:all_abundances}.}
\label{fig:h2o_heatmaps}
\end{figure*}

Figure~\ref{fig:ch4_heatmaps} and Figure~\ref{fig:h2o_heatmaps} at 1.44$\mu$m (a), 1.52$\mu$m (b), and 1.63$\mu$m (c) illustrate the SNR required for a strong \ce{CO2} detection. Figure~\ref{fig:ch4_heatmaps} and Figure~\ref{fig:h2o_heatmaps} at 1.44$\mu$m (d), 1.52$\mu$m (e), and 1.63$\mu$m (f) illustrate the SNR required for a strong \ce{CH4} and \ce{H2O} detection, respectively. The darker the purple, the higher the SNR, and the lighter the pink, the lower the SNR. These heat maps show how \ce{CO2} and \ce{CH4} (Figure~\ref{fig:ch4_heatmaps}) and \ce{CO2} and \ce{H2O} (Figure~\ref{fig:h2o_heatmaps}) affect the detectability of each other. 

Figure~\ref{fig:ch4_heatmaps}a, b and c illustrate the SNR required to achieve a strong \ce{CO2} detection at varying abundances of \ce{CO2} and \ce{CH4}. In Figure~\ref{fig:ch4_heatmaps}a at 1.44$\mu$m, high SNRs (14-17) are required for a strong \ce{CO2} detection with a Venus-like \ce{CO2} abundance at all \ce{CH4} abundances. In Figure~\ref{fig:ch4_heatmaps}b at 1.52$\mu$m, low-high SNRs (9-16) are required for a strong \ce{CO2} detection at a Venus-like \ce{CO2} abundance across all \ce{CH4} abundances and at an Archean-like \ce{CO2} abundance at all except the highest \ce{CH4} abundance. Similarly, in Figure~\ref{fig:ch4_heatmaps}c at 1.63$\mu$m, a strong \ce{CO2} detection is achieved at mid-high SNRs (7-20) for Venus-like and Archean-like abundances of \ce{CO2}. Additionally, a strong \ce{CO2} detection is achievable at high SNRs (18-20) at a Proterozoic-like \ce{CO2} abundance at the two lowest \ce{CH4} abundances. 

Figure~\ref{fig:ch4_heatmaps}d, e and f illustrate the SNR required to achieve a strong \ce{CH4} detection at varying abundances of \ce{CO2} and \ce{CH4}. In Figure ~\ref{fig:ch4_heatmaps}d at 1.44$\mu$m, the two highest \ce{CH4} abundances require high SNRs (17-20) to strongly detect \ce{CH4} at all \ce{CO2} abundances except with a Venus-like \ce{CO2} and second highest \ce{CH4} abundance. In Figure~\ref{fig:ch4_heatmaps}e, we see a significant decrease in SNR requirements at 1.52$\mu$m: low-mid SNRs (4-13) achieve a strong \ce{CH4} detection for the three highest \ce{CH4} abundances at all \ce{CO2} abundances, while medium SNRs are required for a \ce{CH4} abundance of 3$\times10^{-4}$ VMR with \ce{CO2} abundances equal to or lower than 1$\times10^{-1}$ VMR, and high SNRs (19-20) are required for a strong \ce{CH4} detection at the lowest \ce{CH4} abundance and the two lowest \ce{CO2} abundances. In Figure~\ref{fig:ch4_heatmaps}f at 1.63$\mu$m, low-mid SNRs (4-11) achieve a strong \ce{CH4} detection equal to or lower than a \ce{CH4} abundance of 3$\times10^{-4}$ VMR for all \ce{CO2} abundances with the exception of the Venus-like \ce{CO2} and 3$\times10^{-4}$ VMR \ce{CH4} abundance which requires an SNR$=$18. Mid-high SNRs (11-19) achieve a strong \ce{CH4} detection at the lowest \ce{CH4} abundance for all except a Venus-like \ce{CO2} abundance. 

Figure~\ref{fig:h2o_heatmaps}a, b and c illustrate the SNR required to achieve a strong \ce{CO2} detection at varying abundances of \ce{CO2} and \ce{H2O}. In Figure~\ref{fig:h2o_heatmaps}a at 1.44$\mu$m, mid-high SNRs (8-20) are required for a strong \ce{CO2} detection at Venus-like, Archean-like, and Proterozoic-like \ce{CO2} abundances. We see large increases in required SNR as the \ce{CO2} and \ce{H2O} abundances decrease. In Figure~\ref{fig:h2o_heatmaps}b at 1.52$\mu$m, strong \ce{CO2} detections are achievable at low-mid SNRs (6-9) at Venus-like and Archean-like \ce{CO2} abundances and mid-high SNRs (10-16) at a Proterozoic-like \ce{CO2} abundance across all \ce{H2O} abundances. In Figure~\ref{fig:h2o_heatmaps}c at 1.63$\mu$m, a strong \ce{CO2} detection is achievable at low-mid SNRs (5-8) at Venus-like and Archean-like \ce{CO2} abundances and mid-high SNRs (12-15) at a Proterozoic-like \ce{CO2} abundance at all \ce{H2O} abundances. 

Figure~\ref{fig:h2o_heatmaps}d, e and f illustrate the SNR required to achieve a strong \ce{H2O} detection at varying abundances of \ce{CO2} and \ce{H2O}. In Figure~\ref{fig:h2o_heatmaps}d at 1.44$\mu$m, mid-high SNRs (11-18) are required to achieve a strong \ce{H2O} detection at all \ce{CO2} and \ce{H2O} abundances, except at a Venus-like \ce{CO2} abundance and the two lowest \ce{H2O} abundances which require SNRs$>20$. In Figure~\ref{fig:h2o_heatmaps}e at 1.52$\mu$m, low-mid SNRs (4-13) are required to achieve a strong \ce{H2O} detection for all \ce{H2O} abundances across all \ce{CO2} abundances except for a Venus-like \ce{CO2} abundance combined with the lowest \ce{H2O} abundance which requires an SNR$>$20 and the second lowest \ce{H2O} abundance which requires an SNR$=$17. In Figure~\ref{fig:h2o_heatmaps}f at 1.63$\mu$m, low-mid SNRs (5-12) are required for the three highest \ce{H2O} abundances at all \ce{CO2} abundances to achieve a strong \ce{H2O} detection. Mid-high SNRs (13-20) are required for a strong \ce{H2O} detection at the second lowest \ce{H2O} abundance at all \ce{CO2} abundances. No strong \ce{H2O} detections are achievable below an SNR of 20 at the lowest \ce{H2O} abundance. 

To optimize observing and characterizing rocky exoplanets with the HWO coronagraph, the goal is to be sensitive to the largest range of \ce{CO2}, \ce{CH4}, and \ce{H2O} abundances as possible and to achieve strong \ce{CO2} detections at the lowest SNRs. When looking at Figure~\ref{fig:ch4_heatmaps} and Figure~\ref{fig:h2o_heatmaps} as a whole, it is evident that bandpass centers of 1.52$\mu$m and 1.63$\mu$m achieve the most strong \ce{CO2}, \ce{CH4}, and \ce{H2O} detections at the most diverse atmospheric compositions and the lowest SNRs. From previous BARBIE studies, we know that both \ce{H2O} and \ce{CH4} can be strongly detected at shorter wavelengths. Although it would be optimal to observe all three molecules at one wavelength, our main concern lies with the ability to strongly detect \ce{CO2}. Thus, we can focus on Figure~\ref{fig:ch4_heatmaps}b and c and Figure~\ref{fig:h2o_heatmaps}b and c and analyze their differences.

The main difference between Figures~\ref{fig:ch4_heatmaps}b and c are the SNRs required to strongly detect \ce{CO2} at an Archean-like and Proterozoic-like \ce{CO2} abundance: Figure~\ref{fig:ch4_heatmaps}b at 1.52$\mu$m can achieve strong \ce{CO2} detections at lower SNRs at an Archean-like \ce{CO2} abundance. However, Figure~\ref{fig:ch4_heatmaps}c at 1.63$\mu$m, can achieve a strong \ce{CO2} detection at a Proterozoic-like \ce{CO2} abundance at the two lowest \ce{CH4} abundances at very high SNRs (18-20). 

Figures~\ref{fig:h2o_heatmaps}b and c are very similar; Figure~\ref{fig:h2o_heatmaps}c at 1.63$\mu$m requires a slightly lower SNR (1-3 difference) at every \ce{CO2}-\ce{H2O} combination to obtain a strong \ce{CO2} detection compared to Figure~\ref{fig:h2o_heatmaps}b at 1.52$\mu$m. 

However, we cannot make this decision by considering \ce{CO2} detections alone. To determine a long-wavelength cut-off for the HWO coronagraph, we determined that there must be a balance between the detections we can achieve - pushing us to go to longer wavelengths - and the cost of cooling the telescope - pushing us to go to shorter wavelengths. Figures~\ref{fig:ch4_heatmaps} and ~\ref{fig:h2o_heatmaps} reflect the detectability of \ce{CO2} at bandpass centers of 1.52$\mu$m and 1.63$\mu$m. Thus, using a 20\% bandpass width, a bandpass center of 1.52$\mu$m ranges from 1.396-1.679$\mu$m and a bandpass center of 1.63$\mu$m ranges from 1.499-1.803$\mu$m. Since the detections for both bandpasses are roughly equivalent, we find that a long-wavelength cut-off of 1.68$\mu$m would be optimal in detecting \ce{CO2} across various planetary archetypes and SNRs while requiring the least challenging telescope thermal requirements.

\subsubsection{Four Planetary Archetypes}

\begin{figure*}
\centering
\gridline{\fig{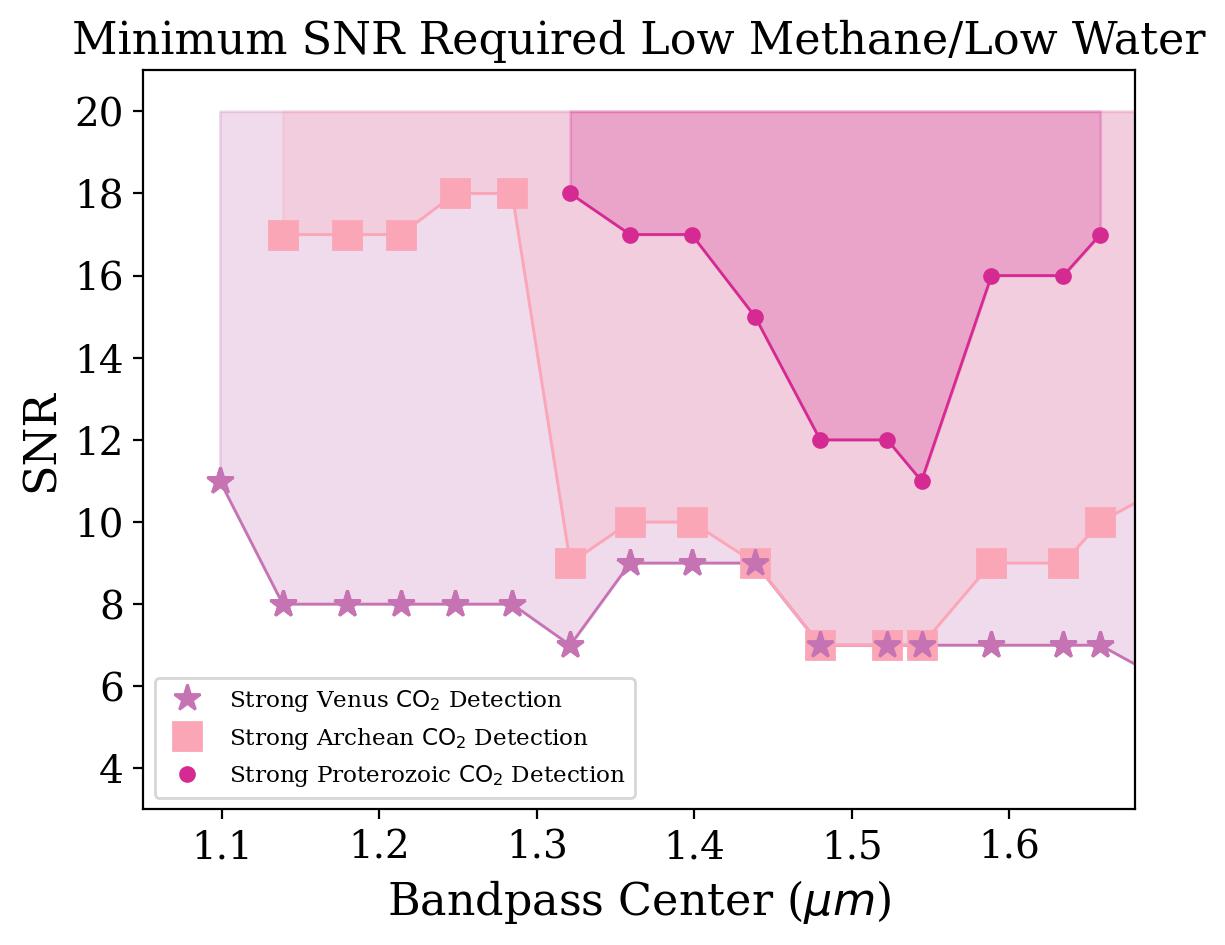}{0.48\textwidth}{\vspace{-0.2cm}(a) Detectability of \ce{CO2} with low abundances of \ce{CH4} and \ce{H2O}}
    \fig{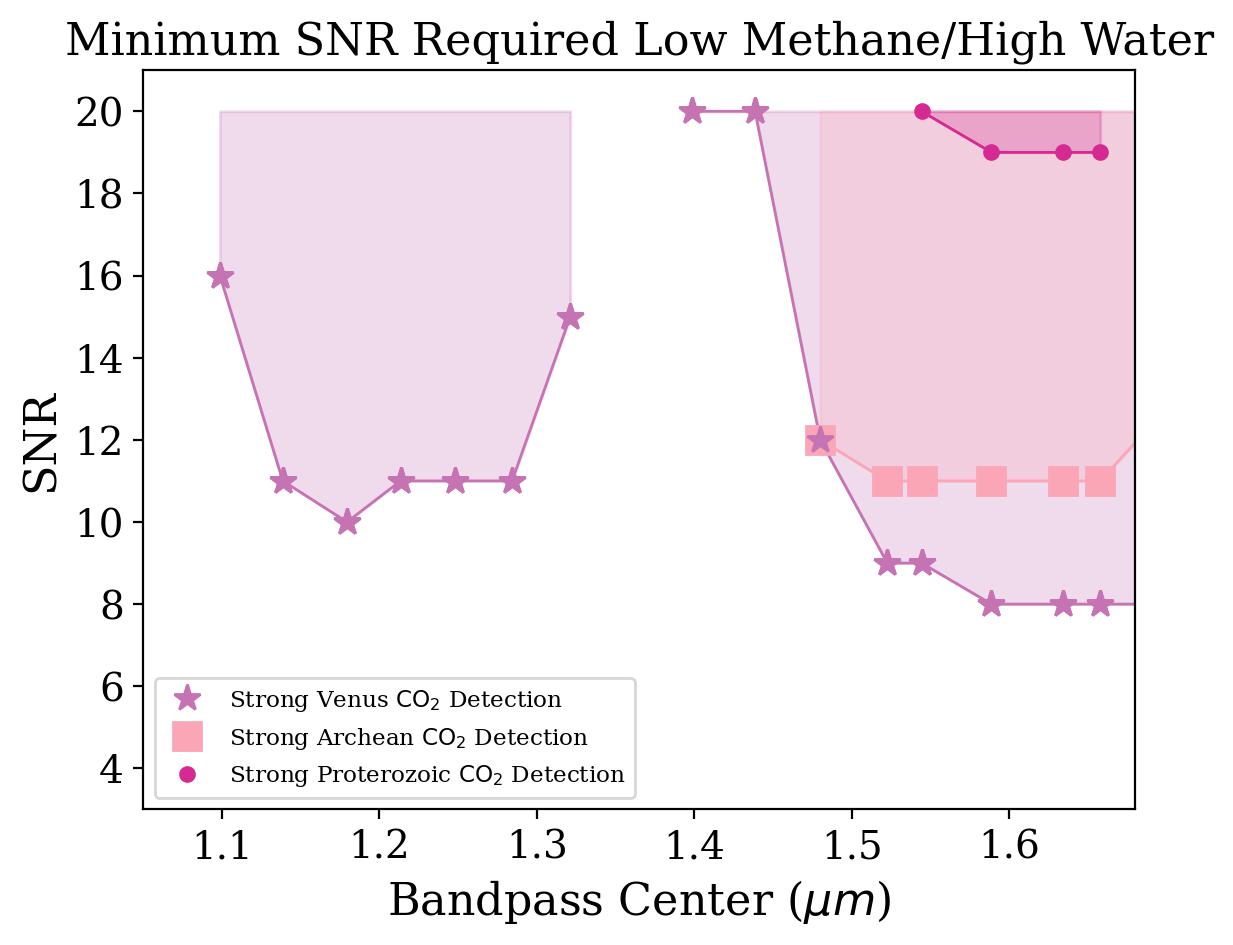}{0.48\textwidth}{\vspace{-0.2cm}(b) Detectability of \ce{CO2} with a low abundance of \ce{CH4} and high abundance of \ce{H2O}}}
\gridline{\fig{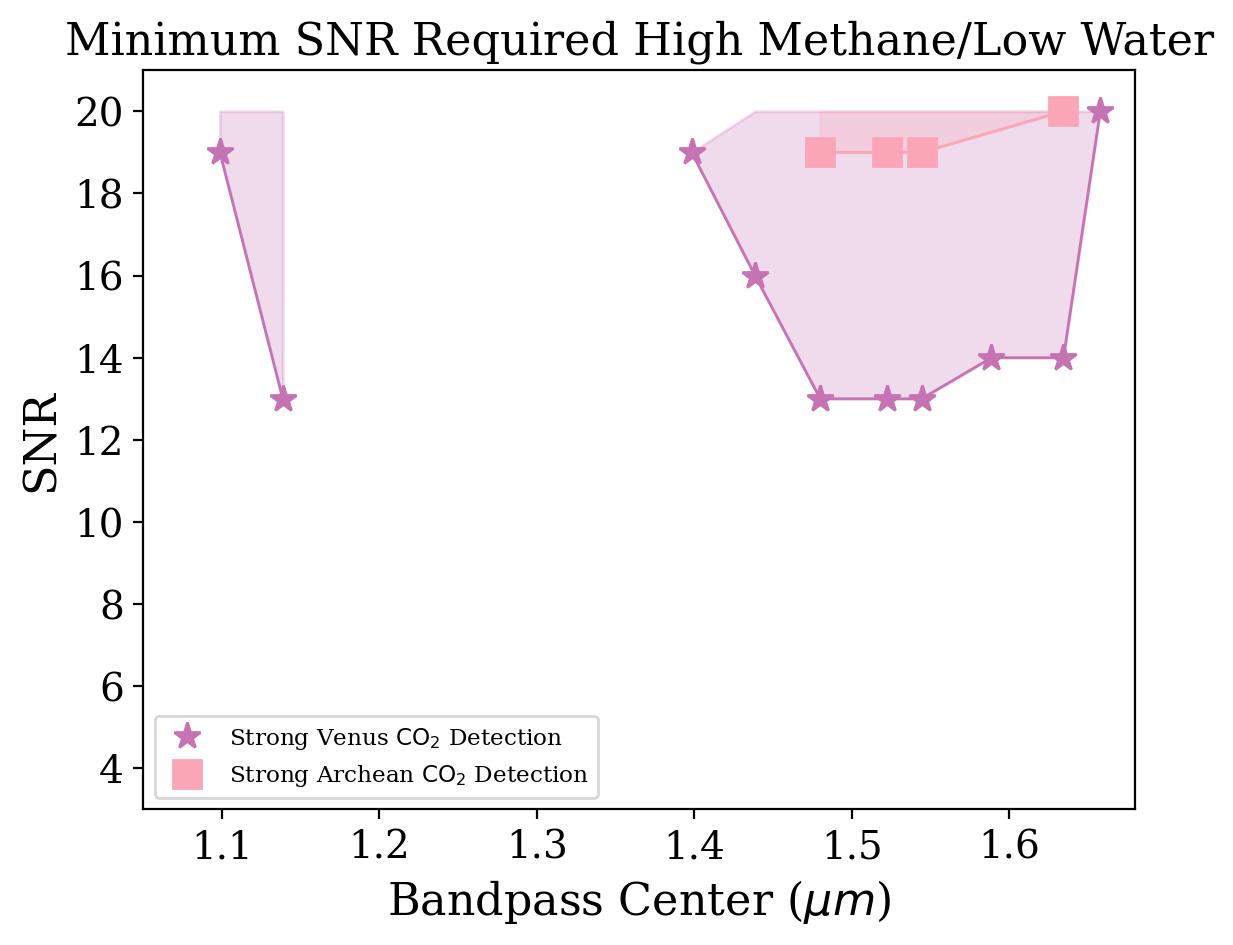}{0.48\textwidth}{\vspace{-0.2cm}(c) Detectability of \ce{CO2} with a high abundance of \ce{H2O} and low abundance of \ce{CH4}}
    \fig{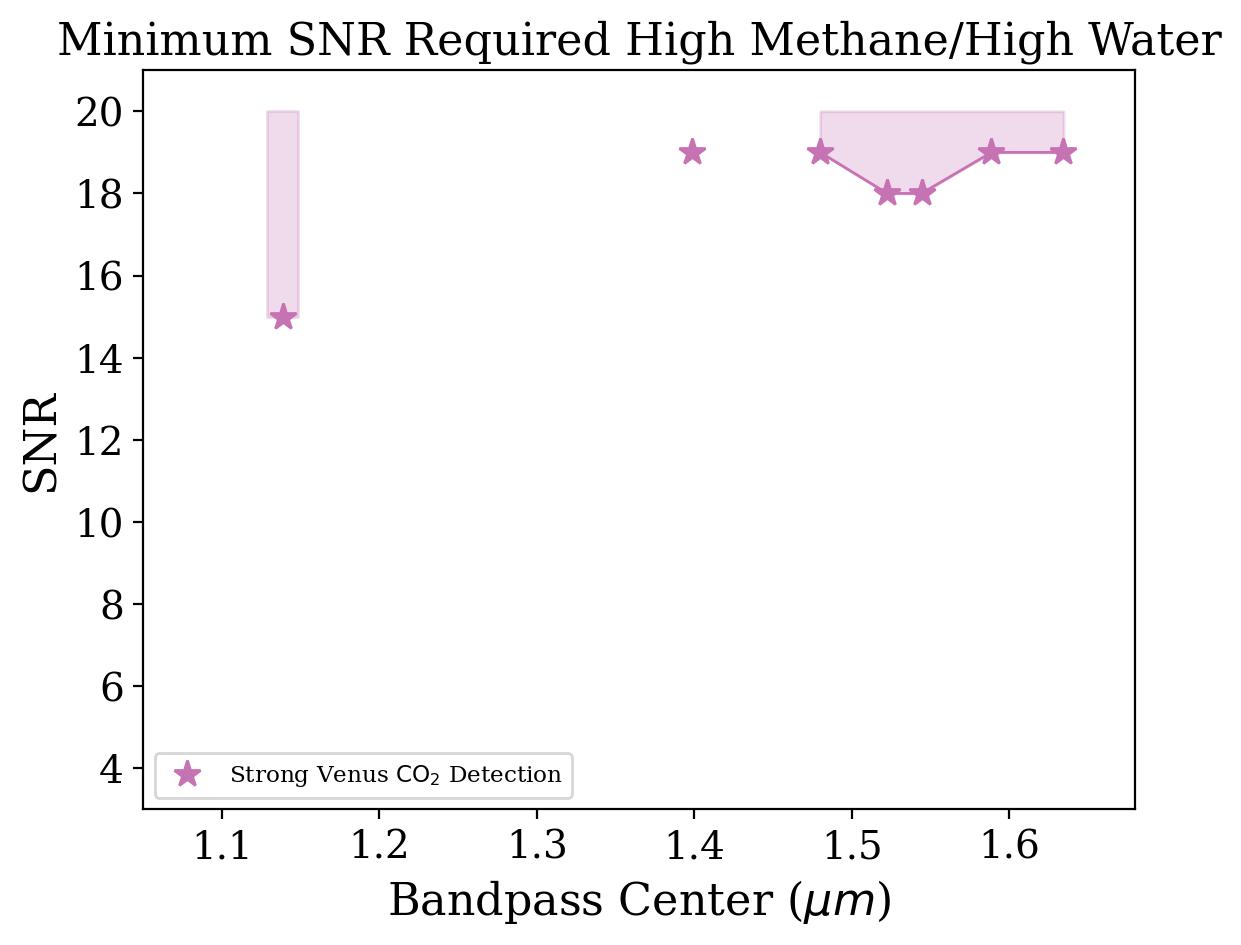}{0.48\textwidth}{\vspace{-0.2cm}(d) Detectability of \ce{CO2} with high abundances of \ce{CO2} and \ce{CH4}}}
\caption{Minimum SNR required to achieve a strong \ce{CO2} detection at each bandpass center for four varying planetary archetypes: (a) low \ce{CH4} \& low \ce{H2O}, (b) low \ce{CH4} \& high \ce{H2O}, (c) high \ce{CH4} \& low \ce{H2O}, and (d) high \ce{CH4} \& high \ce{H2O}. Bandpass center is plotted on the x-axis spanning 1.05-1.68$\mu$m and SNR is plotted on the y-axis. Varying abundances are plotted in different colors and shapes: Venus-like in light purple stars, Archean-like in light pink squares, and Proterozoic-like in dark pink circles. The relative shading corresponds to any area \ce{CO2} is strongly detected and matches colors with the respective \ce{CO2} abundance.}
\label{fig:varying_min_snr}
\end{figure*}

With this optimal long-wavelength cut-off of 1.68$\mu$m for the HWO coronagraph, we can look at how different planetary archetypes would perform. Figure~\ref{fig:varying_min_snr}a, b, c \& d show the minimum SNR required to achieve a strong \ce{CO2} detection at varying combinations of \ce{CH4} and \ce{H2O} abundances: low/low, low/high, high/low \& high/high. A low abundance of \ce{CH4} and \ce{H2O} are both 1$\times10^{-4}$ VMR and a high abundance of \ce{CH4} and \ce{H2O} are 7$\times10^{-4}$ VMR and 1$\times10^{-2}$ VMR, respectively. Strong \ce{CO2} detections are plotted at varying \ce{CO2} abundances: Venus-like in light purple stars, Archean-like in light pink squares, and Proterozoic-like in dark pink circles. The shaded regions indicate areas of strong \ce{CO2} detections for the respective \ce{CO2} abundances.


Comparing the two extremes of low/low in Figure~\ref{fig:varying_min_snr}a and high/high in Figure~\ref{fig:varying_min_snr}d, we see that lower abundances of \ce{CH4} and \ce{H2O} result in more strong \ce{CO2} detections across \ce{CO2} abundance at lower SNRs. This confirms our hypothesis that entangled spectral features can cause degeneracy between molecules. This also informs us that we shouldn't prioritize a planetary archetype similar to Figure~\ref{fig:varying_min_snr}d which achieves strong detections only for a Venus-like \ce{CO2} abundance at high SNRs over a shorter wavelength range compared to Figure~\ref{fig:varying_min_snr}a, b \& c because we wouldn't be able to constrain the planet's atmospheric composition to great accuracy for varying levels of \ce{CO2}. 

Additionally, it is evident that high levels of methane negatively impact the detectability of \ce{CO2} more significantly than a high \ce{H2O} abundance. We know this because \ce{CO2} achieves fewer strong detections in Figure~\ref{fig:varying_min_snr}c with high methane and low water compared to Figure~\ref{fig:varying_min_snr}d with high water and low methane. Thus, we must be vigilant particularly when looking at planetary archetypes with high levels of methane because although we may not achieve a strong \ce{CO2} detection, it does not mean it is not present.

Figure~\ref{fig:varying_min_snr} exemplifies how the complexity of overlapping and saturated features lead to spectral confusion, illustrating the intricacy of \ce{CO2} detections. As spectral features saturate out, there are regions where there are no absorption. In addition to overlapping spectral features, this causes difficulties in differentiating molecules and their spectral features which could lead to confusion in which molecule it is strongly - or not strongly - detecting, impacting our results which is why we see some less intuitive points in our plots. For example, in Figure~\ref{fig:varying_min_snr}a, we see that Venus-like and Archean-like \ce{CO2} abundances achieve strong \ce{CO2} detections at the same SNRs between 1.48-1.54$\mu$m which is interesting considering the vast difference between their VMRs (9.6$\times10^{-1}$ vs 1$\times10^{-1}$). However, if we look at the spectral features in this wavelength range, we are at the edge of a prominent \ce{CO2} feature which could explain why it strongly detects vastly different \ce{CO2} abundances at the same SNR. Comparing this to the middle of the \ce{CO2} spectral feature between 1.59-1.63$\mu$m, Figure~\ref{fig:varying_min_snr}a shows that a Venus-like \ce{CO2} abundance achieves strong \ce{CO2} detections at lower SNRs compared to an Archean-like \ce{CO2} abundance, which intuitively makes sense because a higher abundance correlates to stronger features. Thus, we must consider the spectral complexities when looking at plots like Figure~\ref{fig:varying_min_snr}.


In all four planetary archetypes, a Venus-like \ce{CO2} abundance naturally achieves the most strong detections across our wavelength regime: as low as 1.1$\mu$m in Figures~\ref{fig:varying_min_snr}a, b and c and 1.14$\mu$m in Figure~\ref{fig:varying_min_snr}d. With this in mind, we conceive two different strategies for an initial survey of planetary properties: prioritizing looking for habitability with lower \ce{CO2} abundances such as Archean-like or Proterozoic-like abundances at higher SNRs and longer wavelengths, or looking for uninhabitable atmospheres with ultra-high \ce{CO2} abundances like a Venus-like abundance at shorter wavelengths and lower SNRs. 

With current coronagraphic design conversations, having two parallel channels - one in the visible and one in the NIR - would allow us to confirm or rule out habitability in two ways: (1) observations in the visible, if present, could make a detection of water at wavelengths as short as 0.74$\mu$m \citep{barbie1}, suggesting habitable conditions, or (2) observations in the NIR, if present, could confirm the detection of a Venus-like \ce{CO2} abundance as short as 1.1$\mu$m, suggesting uninhabitable conditions. By having parallel channels, we would not only be able to infer if a planet is habitable, but also have confirmation from the lack of a Venus-like \ce{CO2} abundance.

However, in the case of non-parallel coronagraph channels, we argue that searching for both \ce{H2O} and \ce{CH4} in the visible should be the initial target for spectroscopy, due to their detectability and accessibility in the visible wavelength regime and their implications for life. Additionally, searching for \ce{CO2} in the NIR would take longer than searching for \ce{H2O} and \ce{CH4} in the visible, meaning the priority for detecting \ce{CO2} is secondary when parallel channels are not available.


\section{Conclusions}
\label{sec:end}

In this paper, we explored the conditions necessary to achieve a strong \ce{CO2} detection based on molecular abundance, SNR, and wavelength using the BARBIE methodology and KEN grid set. Using L-KEN, we found that at modern levels of \ce{CO}, there is an insignificant impact on the detectability of \ce{CO2} and no strong \ce{CO} detections are achievable at the highest \ce{CO} abundance. Using B-KEN, we found that with modern levels of \ce{H2O} and \ce{CH4}, only the three highest \ce{CO2} abundances in our test - Venus-like, Archean-like, and Proterozoic-like - result in strong \ce{CO2} detections. We found strong links between SNR, abundance, and wavelength for the case of \ce{CO2}. Generally, stronger detections are more likely at higher SNRs and longer wavelengths. We also found that with increasing abundance, there are more strong detections at lower SNRs and shorter wavelengths. This study also described the detectability relationships of \ce{CO2}-\ce{H2O} and \ce{CO2}-\ce{CH4}. We confirm that there is a strong correlation between these molecules and their impact on detectability of each other at NIR wavelengths. In particular, greater abundances of \ce{H2O} and \ce{CH4} increased the difficulty to detect \ce{CO2}.


Each of our results allow us to better constrain the optimal long-wavelength cut-off for the HWO coronagraph and understand the inter-connectedness between \ce{CO2}, \ce{H2O}, and \ce{CH4}. Figure~\ref{fig:intro}, showing thermal noise due to the blackbody emission of the telescope as a function of telescope temperature and wavelength (top) and \ce{CO2}, \ce{CH4}, \ce{H2O}, \ce{CO}, and \ce{O2} spectral features between 0.8-2.0$\mu$m (bottom), illustrates that longer wavelengths would drastically increase the cost from cooling the telescope due to the increase in noise as the telescope warms and emphasizes the rationale for studying \ce{CO2} due its prominent features at longer wavelengths - specifically looking at the feature that peaks at 1.6$\mu$m. Figure~\ref{fig:heatmaps} showed us that analyzing the wavelength regime between the bandpass centers of 1.52-1.63$\mu$m would result in the maximum number of strong detections across \ce{CO2} abundance and SNR. Figure~\ref{fig:minimum_snr} showed us that Venus-like, Archean-like, and Proterozoic-like \ce{CO2} abundances can be strongly detected at their respective lowest SNRs starting at 1.59$\mu$m and proved that it is unnecessary to go longer than 1.7$\mu$m because it does not provide additional valuable information. Figure~\ref{fig:detectability} emphasized that there are outside factors besides \ce{CO2} abundance and SNR that could impact \ce{CO2} detectability which led us to simulate tests varying \ce{CH4} and \ce{H2O} abundances. Figure~\ref{fig:1.78_heatmaps} confirmed that an increased abundance of \ce{CH4} resulted in less strong \ce{CO2} detections and a decreased abundance of \ce{H2O} resulted in more strong \ce{CO2} detections. Because of this, we tested a variety of different \ce{CH4} and \ce{H2O} abundances as seen in Figure~\ref{fig:ch4_heatmaps} and Figure~\ref{fig:h2o_heatmaps}, respectively. These plots demonstrated that 1.52$\mu$m was the optimal bandpass center that provided the most strong \ce{CO2} detections across all abundances with the lowest SNRs and reasonable cost. In addition, although \ce{CH4} and \ce{H2O} can be strongly detected at shorter wavelengths, as seen in previous BARBIE studies, 1.52$\mu$m also provided optimal \ce{CH4} and \ce{H2O} detections. Figure~\ref{fig:varying_min_snr} showing \ce{CO2} detectability at various planetary archetypes emphasized that lower abundances of \ce{CH4} and \ce{H2O} positively impact our ability to strongly detect \ce{CO2}. In addition, it confirmed that a long-wavelength cut-off of 1.68$\mu$m - with a bandpass center of 1.52$\mu$m - is optimal for strongly detecting various abundances of \ce{CO2}. However, detecting a Venus-like \ce{CO2} abundance as low as 1.1$\mu$m could allow us to rule out a habitable atmosphere if present.



By combining all of our results, we conclude that the optimal bandpass center to strongly observe \ce{CO2} is 1.52$\mu$m, meaning that 1.68$\mu$m is the optimal long-wavelength cut-off for the HWO coronagraph to be able to sufficiently achieve strong \ce{CO2} detections at varying \ce{CO2}, \ce{H2O}, and \ce{CH4} abundances at the lowest SNRs. 

The next steps in the BARBIE world are to explore the molecular detection requirements in the UV using the KEN grid set, which will help dictate the technological requirements of HWO in the UV. With the KEN grids, our studies focus on a specific set of biosignatures. We encourage others to study the detectability of other potential biosignatures such as methyl halides (\ce{CH3Cl}, \ce{CH3Br}, and \ce{CH3I}), dimethyl sulfide (DMS), and dimethyl disulfide (DMDS) to confirm whether a 1.68$\mu$m cut-off for the HWO coronagraph will be sufficient for strongly detecting these molecules as well.

\section*{Acknowledgments}
C. H. acknowledges the financial support from SURA and CRESST II. She acknowledges Dr. Avi Mandell and Dr. Natasha Latouf for their support and guidance as they brought her onto the BARBIE team. She also gratefully acknowledges Greta Gerwig, Margot Robbie, Ryan Gosling, Emma Mackey, and Mattel Inc.{\texttrademark} for Barbie (doll, movie, and concept), for which this project is named after. N. L. gratefully acknowledges financial support from a NASA FINESST and an appointment to the NASA Postdoctoral Program (NPP) at the NASA Goddard Space Flight Center, administered by Oak Ridge Associated Universities under contract with NASA. The material is based upon work supported by NASA under award number 80GSFC24M0006. The authors would like to thank the Sellers Exoplanet Environments Collaboration (SEEC) and ExoSpec teams at NASA's Goddard Space Flight Center for their consistent support. These Barbies are astrophysicists!

\bibliographystyle{aasjournal}
\bibliography{main}

\end{document}